\documentclass[]{aastex}

\usepackage{emulateapj5,ifthen}
\shorttitle{Near-IR Properties of Galaxy Clusters}
\shortauthors{Lin, Mohr \& Stanford}

\def\Om{$\Omega_M$\ }

\def\xray{\hbox{X--ray} }
\def\nir{\hbox{NIR} }
\def\nire{\hbox{NIR}}
\def\kev{{\rm\ keV}}
\def\mlr{\hbox{mass--to--light ratio}\ }
\def\mlrs{\hbox{mass--to--light ratios}\ }
\def\mlre{\hbox{mass--to--light ratio}}

\newcommand{\figtype}{EPS}
\def\myputfigure#1#2#3#4#5%
{\vskip#5pt\makebox[0pt]{\hskip#2in
\includegraphics[width=#3\textwidth]{#1}}\vskip#4pt\hfill}
\newenvironment{inlinefigure}{
\def\@captype{figure}
\noindent\begin{minipage}{0.999\linewidth}\begin{center}}
{\end{center}\end{minipage}\smallskip}

\begin{document}

\submitted{Submitted to ApJ December 6, 2002, Accepted March 25, 2003}

\title{Near-IR Properties of Galaxy Clusters:
       Luminosity as a Binding Mass Predictor and the State of Cluster Baryons}

\author{Yen-Ting Lin\altaffilmark{1}, Joseph J. Mohr\altaffilmark{1,2}
and S. A. Stanford\altaffilmark{3,4}}
\altaffiltext{1}{Department of Astronomy, University of Illinois,
Urbana, IL 61801; ylin2@astro.uiuc.edu}
\altaffiltext{2}{Department of Physics, University of Illinois,
Urbana, IL 61801; jmohr@uiuc.edu}
\altaffiltext{3}{Physics Department, University of California at Davis,
Davis, CA 95616; adam@igpp.ucllnl.org}
\altaffiltext{4}{Institute of Geophysics and Planetary Physics, Lawrence 
Livermore National Laboratory, Livermore, CA 94551}

\begin{abstract}

We explore the near-infrared properties of galaxies within 27 galaxy
clusters using data from the Two Micron All Sky Survey (2MASS).  For a
subsample of 13 clusters with available X-ray imaging data, we examine both
the properties of the galaxies and the intracluster medium.  We show
that the $K$-band luminosity is correlated with cluster mass,
providing a binding mass estimate accurate to 45\%.  The \mlr in our
ensemble increases by a factor of $\sim 2$ over the cluster mass range 
($10^{14}M_\odot-10^{15}M_\odot$).  We examine the total baryon 
fraction, showing that it is an increasing function of cluster mass.  Using the 
\mlr of massive clusters, we find that $\Omega_M=0.19\pm0.03$; using the total
baryon fraction we find that $\Omega_M=0.28\pm0.03$, in good agreement with recent cosmic microwave background (CMB) anisotropy constraints.  Differences 
between these two estimates suggest that the $K$-band \mlr in 
massive clusters may be lower than that in the universe by as much as $\sim30\%$.  

We examine the stellar mass fraction, the intracluster medium (ICM) 
mass to stellar mass ratio and the cluster iron 
mass fraction.  The stellar mass fraction decreases by a factor of 1.8
from low to high mass clusters, 
and the ICM to stellar mass ratio increases from 5.9 to 10.4 over the same 
mass range.  Together, these measurements suggest a decrease of star 
formation efficiency with increasing cluster mass and provide 
constraints on models of the thermodynamic history of the intracluster medium.  
The cluster iron mass to total mass ratio is constant and high, 
suggesting that some efficient, and uniform enrichment process may have 
taken place before the bulk of stars in cluster galaxies formed.
\end{abstract}

\keywords{cosmology: observation -- galaxies: clusters, luminosity function
  -- infrared: galaxies}

\section{Introduction}
\label{sec:intro}

Galaxy clusters are central to many cosmological studies.  The local
cluster abundance is a direct measure of the present-day matter power
spectrum amplitude and the dark matter density \citep{swhite93b,viana99,
reiprich02}. The cluster redshift distribution constrains
the growth of density perturbations, and this effect can be used to
constrain the amount and nature of the dark matter and dark energy
\citep{wang98,haiman01,holder01b,levine02,majumdar02}.  The baryon fraction
of the most massive local galaxy clusters has been used to constrain the matter 
density parameter \citep{swhite93a,david95,dwhite95,mohr99,grego01}, and the 
spatial distribution of galaxy clusters has been used to measure the power 
spectrum of density perturbations \citep{bahcall83,miller01,schuecker02}.

The galaxy cluster halo mass is crucial to the interpretation of each of the 
cosmological measurements mentioned above.  The cluster halo mass, which is 
dominated by dark matter, is not a direct observable; a good mass indicator 
is required to bridge the gap between the observational and  theoretical 
realms.  Detailed mass estimators relying on hydrostatic or virial equilibrium 
or on the weak lensing distortions of background galaxies have been developed, 
but generally these estimates require extensive \xray and optical imaging 
and/or spectroscopy.  The large cluster surveys that are being planned and 
carried out to measure the cluster redshift and spatial distribution will 
produce large samples dominated by systems that are near the detection 
threshold.  Deep follow up of each of these systems with an \xray telescope 
or multiobject spectrograph is implausible, and so the analyses of these 
future surveys will rely on the existence of simpler and perhaps less
accurate mass estimators.  Specifically, these surveys will employ scaling 
relations between simple cluster observables like the \xray luminosity or 
the Sunyaev-Zel'dovich effect \citep[SZE;][]{sunyaev70,sunyaev72} luminosity 
and cluster halo mass.  

Observations over the past several 
years have demonstrated that clusters exhibit significant regularity 
\citep{mohr97a,arnaud99,mohr99}, and we now have several locally calibrated 
mass--observable relations.  Among these are the mass--\xray temperature 
relation \citep[$M-T_X$, e.g.][]{finoguenov01}, the mass--luminosity relation 
\citep[$M-L_X$, e.g.][]{reiprich02} and the mass--velocity dispersion relation 
\citep[$M-\sigma_V$, e.g.][]{girardi00}.  Here we use the Two Micron All Sky 
Survey (2MASS) to explore whether the galaxy light itself is a good indicator of
the cluster mass by examining the cluster halo mass-$K$-band light relation.  
Such a relation would be extremely valuable for interpreting any cluster survey 
with near-IR (hereafter NIR) coverage.

In addition to the cosmological studies, there is a range of unresolved 
questions surrounding the star formation history of the universe, and the 
thermodynamic history of the intergalactic and intracluster medium 
(hereafter ICM).  The galaxy \nir light in combination with \xray imaging 
and spectroscopy are well suited to address these issues, because the \xray 
observations provide measures of the halo and ICM masses and the \nir light 
appears to be a good tracer of the total stellar mass.  The effects of dust 
in \nir wavebands are rather small compared to optical and ultraviolet (UV) 
wavebands. The \nir light is less sensitive to the recent star formation 
history in galaxies.  Studies of a large sample of spiral galaxies 
\citep{gavazzi96} indicate that the $H$-band luminosity is proportional to 
the galaxy dynamical mass, implying that $H$-band light can serve as a better 
mass estimator than $U$, $B$ or $V$-band light.  Modeling suggests that the 
$K$--band \mlr can still vary by as much as a factor of two over a range of 
galaxy Hubble type, color, and star formation histories \citep{madau98,
pahre98}, so care must be taken to account for these variations wherever 
possible.

Massive galaxy clusters largely exhibit self-similar scaling relations, but 
that behavior does not hold for less massive clusters and groups 
\citep[e.g.][]{lloyd-davies00,voit02}.  Non-gravitational processes, such as 
feedback due to star formation activity in galaxies or AGN heating may 
significantly alter the structure of the lower mass systems, but important 
questions remain about which processes are important.  By studying the state 
of the baryons in present epoch clusters, one can in principle unravel details 
of their formation and evolution.  For example, the distribution of the ICM 
gas may give clues to the importance of non-gravitational heating 
\citep{cavaliere97,mohr99,bialek01}; the total stellar content within a 
cluster measures the cooled baryon fraction, which in turn describes the 
efficiency of galaxy formation in the clusters \citep{bryan00b}; the 
metallicity of the ICM gas also provides stringent constraints on the star 
formation history of cluster galaxies, as well as on the mechanisms that 
transport the metals from the galaxies into the ICM \citep{arnaud92,
renzini97}. Among the 27 clusters in our sample, we have measured the ICM 
mass for a subsample of 13 systems. This, together with the total stellar 
mass inferred from the \nir luminosity, gives us an opportunity to address 
different preheating, star formation and enrichment scenarios.

This paper is structured as follows. In \S\ref{sec:data} we describe
how we infer the cluster virial mass from \xray temperature, how the
total \nir galaxy luminosity is obtained from the 2MASS data, and how
the ICM mass is measured using \xray data. We present the \nir
mass--luminosity relation, as well as the \mlr for our sample of 
clusters in \S\ref{sec:nirlite}. In the rest of \S\ref{sec:nirlite}
we examine constraints on the density parameter \Om using the cluster
mass--to--light ratio (\S\ref{sec:M2L}) and the total baryon
fraction (\S\ref{sec:totbaryon}).  We compare the results and discuss
the cosmological implications in \S\ref{sec:systematics}. In 
\S\ref{sec:icmstar} we examine the cluster baryon reservoirs in more detail.
The discussion includes the star formation efficiency on cluster mass scales 
(\S\ref{sec:starformation}), the relative distribution of baryons in the
ICM and in galaxies (\S\ref{sec:star2icm}), and the ICM iron abundance
and clues on the enrichment (\S\ref{sec:enrich}). Finally, we present our
conclusions, discuss the robustness of our analysis, the possible systematics, 
and prospects for further investigation in \S\ref{sec:err} \& \S\ref{sec:sum}.

Throughout the paper we assume the density parameters for the matter and the
cosmological constant to be $\Omega_M = 0.3$, $\Omega_\Lambda = 0.7$, 
respectively, and the Hubble parameter to be
$H_0=70\,h_{70}$~km~s$^{-1}$~Mpc$^{-1}$.

\section{Joint NIR and X--ray Analysis}
\label{sec:data}

We use the observed \xray mass--temperature relation together with published 
\xray emission weighted mean temperatures, 2MASS second incremental release \nir
data, and \xray imaging data to study trends in the \nir and \xray properties of
galaxy clusters.  Our cluster sample is based on existing cluster samples
(\citealt{david93,mohr99}, hereafter MME; \citealt{reiprich02}) and is
limited by the incomplete sky coverage in the 2MASS second release
($\sim 47\%$ of the sky).  We return to the issue of sample selection 
and possible effects on our results in \S \ref{sec:lf} \& \ref{sec:err}.
The relevant cluster parameters (\xray emission center, redshift $z$, \xray 
emission weighted temperature $T_X$) are gathered from the above references and 
others \citep{edge91,dwhite00}.  We only consider the clusters that have
reliable $T_X$ measurements and are reasonably beyond the galactic plane 
($|b|>10^\circ$). The 2MASS second release contains 27
clusters for which luminosity functions can be estimated in the method
described below. Among these 27 clusters, the ICM masses $M_{ICM}$ of
13 were measured by MME, based on an analysis of ROSAT PSPC
observations.  We shall refer to these as the MME subsample.
A list of our cluster sample appears in Table 1.  In short, our
cluster sample is at low redshift ($0.016 \lesssim z \lesssim 0.09$ with
mean redshift $\sim 0.05$), and spans a range of \xray emission
weighted temperature ($2.1 \le T_X \le 9.1$) that corresponds to about
an order of magnitude in cluster binding mass ($0.8 - 9 \times
10^{14}\,h_{70}^{-1}\, M_\odot$).

\subsection{X--ray Estimate of Cluster Mass}
\label{sec:mt}

We use an observed \xray $M-T_X$ relation to calculate the binding mass 
for each cluster. \citet{finoguenov01}   
provide several $M-T_X$ relations that arise from different
subsets of a cluster ensemble.  We use the relation
obtained by fitting to the clusters with ASCA temperature profiles that are more
massive than $3.57\times 10^{13} \,h_{70}^{-1}\, M_\odot$; all our 
clusters lie in this mass range. To be definite, we use
\begin{equation}
\label{eq:MT}
M_{500} = 2.55^{+0.29}_{-0.25}\,10^{13} {M_\odot\over h_{70}}\,
    \left( {T_X \over 1\kev} \right)^{1.58^{+0.06}_{-0.07}}
    \label{eq:m-t}
\end{equation}
where $M_{500}$ is the mass enclosed by $r_{500}$, within which the mean 
overdensity is $500$ times of the critical density of the universe $\rho_c$. 
Because our clusters are all low redshift, we neglect the effects of cosmic 
density evolution. 

Within our fiducial cosmological model ($\Omega_M=0.3$ and $\Omega_\Lambda=0.7$)
we calculate the cluster radius $r_{500}$ and its angular extent $\theta_{500} 
\equiv r_{500}/d_A$, where $d_A$ is the angular diameter distance. We analyze 
the galaxy and ICM properties within this region of the cluster.  At the
low redshift range spanned by our cluster sample the sensitivity to
cosmology in the angular diameter distance is unimportant.

\subsection{NIR Luminosity from 2MASS Data}
\label{sec:2mass}

Knowing the cluster angular extent, we search the 2MASS extended source catalog 
to identify cluster member galaxies. Among the three bands ($J$, $H$ and
$K_s$) 2MASS provides, $K_s$ has been studied most extensively. For
this reason we shall only consider the $K_s$-band (hereafter denoted
as $K$-band for simplicity) in this paper. We use the ``default''
$K_s$ magnitudes. The official 2MASS second
release completeness limit for the default $K_s$-band is $K_{lim} = 13.5$
\footnote{http://www.ipac.caltech.edu/2mass/releases/second/doc/explsup.html}.
However, after visual inspection of the magnitude distribution
of the galaxies, we choose $K_{lim} = 13.3$ as our completeness limit.

\begin{inlinefigure}
   \ifthenelse{\equal{\figtype}{EPS}}{
   \begin{center}
   \epsfxsize=8.cm
   \begin{minipage}{\epsfxsize}\epsffile{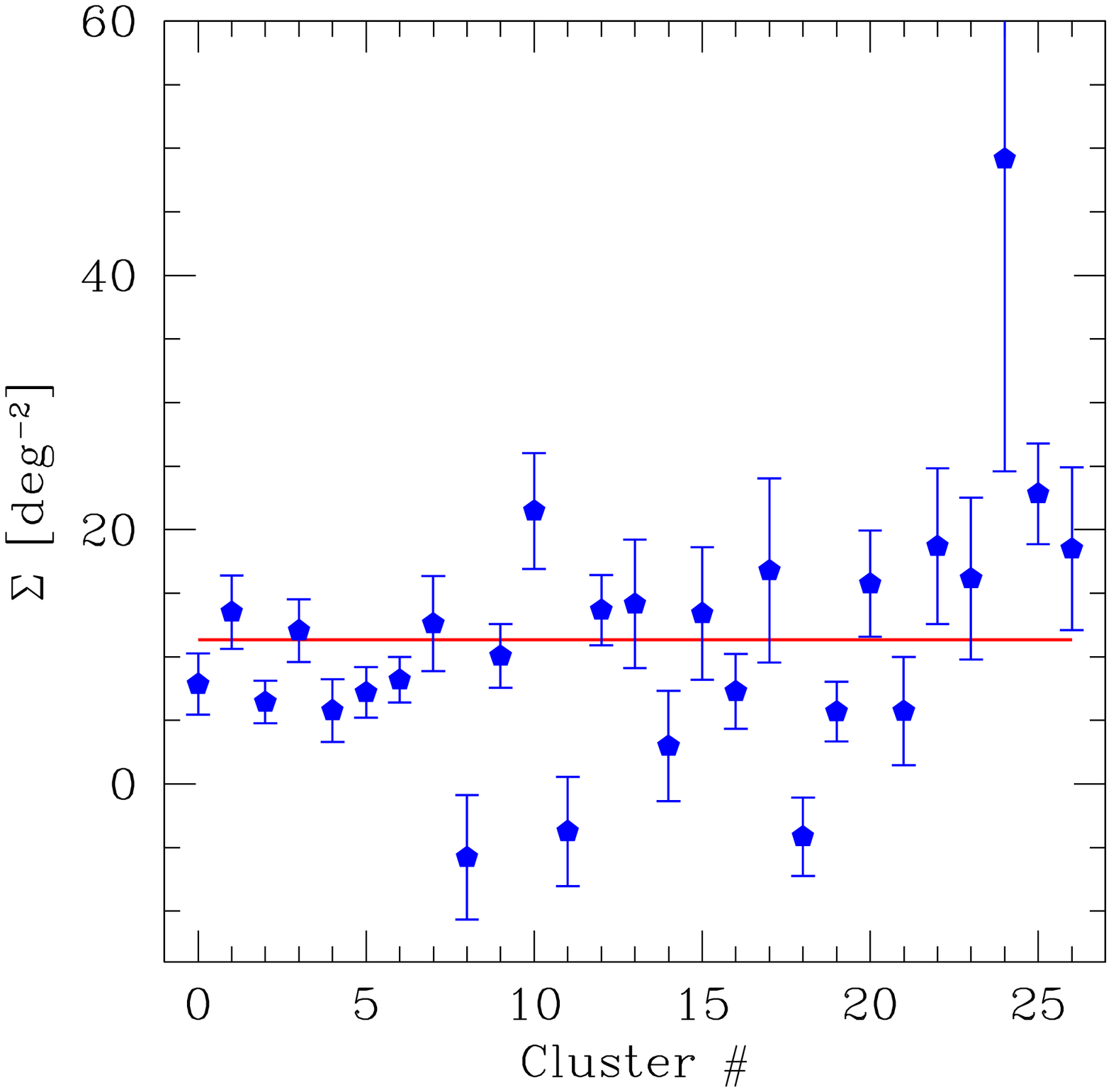}\end{minipage}
   \end{center}}
   {\myputfigure{f1.pdf}{0.0}{1.0}{-70}{-20}}
   \figcaption{\label{fig:bgn}
	A comparison between the background galaxy surface density
	as obtained using the statistical method ($\Sigma_{stat}$, solid
        line), and the ``annulus'' method ($\Sigma_{bgn}$, solid points, see
        text)
        }
\end{inlinefigure}

\subsubsection{Background Correction}
\label{sec:bgn}

To correct both the number counts and fluxes from
background (and foreground; hereafter we refer to all non-cluster galaxies as background galaxies) galaxies in estimating the cluster \nir luminosity, we use 
the observed $K$-band galaxy brightness distribution $d\,\Sigma_{stat}/dK$ 
\citep{kochanek01}. Integrating this quantity with respect to magnitude to our 
adopted 2MASS completeness limit gives the statistical galaxy surface density
$\Sigma_{stat}$. This, multiplied by the cluster angular area
$\pi \theta_{500}^2$, gives the expected number of background 
galaxies.  The expected flux from the background galaxies is
$F_{stat} = \pi \theta_{500}^2 \int_{-\infty}^{K_{lim}}\,
(d\,\Sigma_{stat}/dK)\,F\,dK$, where $F$ is the galaxy flux.

As a check of this statistical method, we also perform an ``annulus'' background
correction; that is, for an assumed galaxy distribution profile with
unknown normalization and a unknown constant background surface density,
for each cluster we measure 
the number of galaxies within $\theta_{500}$ ($N_{cen}$) and
within an annulus outside the cluster virial region ($N_{ann}$), whose
inner and outer radii are $1.5\,\theta_{200}$ and $3\,\theta_{200}$,
respectively, where $\theta_{200} \equiv r_{200}/d_A$. 
We convert between $r_{200}$
and $r_{500}$ using the ``universal'' dark matter distribution profile
found in N-body simulations \citep[][NFW]{navarro97}; we use
a fixed ``concentration'' parameter $c=5$, and in that case $r_{200} =
1.51\,r_{500}$.

Background and cluster galaxies lie in both the central and annular
regions.  Using the observables $N_{cen}$ and 
$N_{ann}$ and the NFW model for the distribution of galaxies, we estimate 
the normalizations of the cluster galaxy distribution 
and the uniform background galaxy surface density
$\Sigma_{bgn}$. The number of cluster galaxies within $\theta_{500}$
is then $N_{cen}-\Sigma_{bgn}\, \pi \theta_{500}^2$.  For some
clusters the annulus is only partially surveyed in the 2MASS
second release, and we correct for this partial coverage. In some cases
there are obvious galaxy clusterings (from visual inspection) in the
annulus region; we avoid this possible contamination of the background
annulus by other clusters or groups by excluding the area occupied
by the clustering in the annulus.

Fig~\ref{fig:bgn} contains a comparison of $\Sigma_{stat}$ determined from the 
statistical approach and $\Sigma_{bgn}$ as determined from the ``annulus'' 
method. There is general agreements between $\Sigma_{stat}$ and $\Sigma_{bgn}$, indicating that local and statistical background corrections lead to the same results.
For the analyses that follow, we adopt the statistical background surface
density in making background corrections to the cluster galaxies.
Note that, for 3 clusters,  $\Sigma_{bgn}$ determined from the ``annulus'' 
method is negative, which may either be due to the correction to the 
partially surveyed annulus region, or be due to a significant mismatch between 
the actual galaxy distribution in these clusters and the NFW model we are using.
We have checked all the scaling relations presented in the following analysis 
using the local background estimate, and the changes to slope and normalization 
all lie within the 1$\sigma$ uncertainties.

\begin{table*}[htb]
\begin{center}
\caption{Basic Descriptions of the Clusters}
{\scriptsize
\begin{tabular}{lclrccclcc}
\tableline \tableline
  (1) & (2)  & (3)   &  (4)       & (5)                    & (6)   & (7)                             & (8)                       & (9)          & (10) \\  
      &      & $T_X$ &            &  $\phi_*$              & $M_*-5\log h_{70}$ &  $L_{500}$                      &  $\Upsilon_{500}$         & $f_{star}$   & $f_{ICM}$ \\
Name  & $z$  & (keV) &  $N_{gal}$ &($h_{70}^3\,$Mpc$^{-3}$)& (mag) &($h_{70}^{-2}\,10^{11} L_\odot$) & ($h_{70}\,\Upsilon_\odot$)& ($10^{-2}$)  & ($10^{-2}$) \\
\tableline
A85 & 0.0556 & $6.1^{0.12}_{0.12}$ & 27 & $5.26^{1.81}_{1.81}$ & $-24.71 \pm {0.18}$ & $7.97 \pm 0.03$ & $55^{14}_{14}$ & $1.41^{0.46}_{0.46}$ & $14.92 \pm {0.39}$ \\
A119 & 0.0440 & $5.8^{0.36}_{0.36}$ & 39 & $3.80^{1.28}_{1.28}$ & $-25.24 \pm {0.20}$ & $8.32 \pm 0.05$ & $49^{13}_{13}$ & $1.58^{0.53}_{0.53}$ & $12.03 \pm {0.75}$ \\
A262 & 0.0161 & $2.41^{0.03}_{0.03}$ & 41 & $7.31^{2.37}_{2.37}$ & $-24.69 \pm {0.27}$ & $2.51 \pm 0.04$ & $40^{10}_{10}$ & $1.76^{0.57}_{0.57}$ & $10.17 \pm {0.39}$ \\
A496 & 0.0328 & $3.91^{0.04}_{0.04}$ & 40 & $7.41^{2.36}_{2.36}$ & $-24.59 \pm {0.20}$ & $4.96 \pm 0.03$ & $44^{11}_{11}$ & $1.71^{0.55}_{0.55}$ & $13.47 \pm {0.32}$ \\
A644 & 0.0704 & $6.59^{0.1}_{0.1}$ & 17 & $5.57^{2.15}_{2.15}$ & $-24.68 \pm {0.19}$ & $8.45 \pm 0.03$ & $59^{15}_{15}$ & $1.33^{0.44}_{0.44}$ & $13.04 \pm {0.47}$ \\
A754 & 0.0528 & $8.5^{0.3}_{0.3}$ & 44 & $4.04^{1.30}_{1.30}$ & $-24.86 \pm {0.16}$ & $10.99 \pm 0.03$ & $68^{17}_{17}$ & $1.17^{0.38}_{0.38}$ & $13.90 \pm {1.50}$ \\
A1367 & 0.0216 & $3.5^{0.11}_{0.11}$ & 59 & $7.25^{2.24}_{2.24}$ & $-24.77 \pm {0.20}$ & $4.42 \pm 0.03$ & $41^{10}_{10}$ & $1.79^{0.58}_{0.58}$ & $11.85 \pm {0.45}$ \\
A1651 & 0.0850 & $6.3^{0.31}_{0.31}$ & 8 & $17.55^{8.81}_{8.81}$ & $-24.06 \pm {0.17}$ & $13.89 \pm 0.01$ & $33^{10}_{10}$ & $2.33^{0.88}_{0.88}$ & $13.69 \pm {0.70}$ \\
A2255 & 0.0800 & $6.87^{0.2}_{0.2}$ & 16 & $3.36^{1.30}_{1.30}$ & $-25.30 \pm {0.21}$ & $9.81 \pm 0.06$ & $54^{14}_{14}$ & $1.45^{0.48}_{0.48}$ & $13.14 \pm {2.60}$ \\
A2319 & 0.0564 & $9.12^{0.09}_{0.09}$ & 61 & $5.95^{1.80}_{1.80}$ & $-24.83 \pm {0.12}$ & $17.30 \pm 0.02$ & $48^{12}_{12}$ & $1.65^{0.53}_{0.53}$ & $16.69 \pm {0.52}$ \\
A3266 & 0.0594 & $6.2^{0.5}_{0.4}$ & 30 & $4.92^{1.78}_{1.73}$ & $-25.03 \pm {0.18}$ & $9.94 \pm 0.04$ & $45^{13}_{12}$ & $1.71^{0.59}_{0.58}$ & $16.66 \pm {1.31}$ \\
A3558 & 0.0480 & $5.7^{0.12}_{0.12}$ & 61 & $7.59^{2.30}_{2.30}$ & $-25.09 \pm {0.14}$ & $13.91 \pm 0.03$ & $28^{7}_{7}$ & $2.72^{0.88}_{0.88}$ & $15.83 \pm {0.38}$ \\
A4038 & 0.0283 & $3.15^{0.03}_{0.03}$ & 32 & $6.50^{2.15}_{2.15}$ & $-24.68 \pm {0.24}$ & $3.23 \pm 0.04$ & $48^{12}_{12}$ & $1.53^{0.49}_{0.49}$ & $11.64 \pm {0.73}$ \\
A133 & 0.0569 & $3.8^{1.25}_{0.56}$ & 13 & $14.15^{9.92}_{6.94}$ & $-23.95 \pm {0.22}$ & $4.97 \pm 0.02$ & $42^{25}_{15}$ & $1.79^{1.12}_{0.75}$ & -- \\
A548e & 0.0410 & $3.1^{0.1}_{0.1}$ & 31 & $7.71^{2.57}_{2.57}$ & $-25.15 \pm {0.22}$ & $5.72 \pm 0.05$ & $26^{6}_{6}$ & $2.78^{0.90}_{0.90}$ & -- \\
A1185 & 0.0304 & $3.9^{2}_{1.1}$ & 27 & $4.01^{3.79}_{2.41}$ & $-24.81 \pm {0.37}$ & $3.25 \pm 0.06$ & $67^{57}_{34}$ & $1.12^{0.98}_{0.62}$ & -- \\
A1650 & 0.0845 & $6.7^{0.5}_{0.5}$ & 6 & $4.66^{2.60}_{2.60}$ & $-24.55 \pm {0.26}$ & $6.74 \pm 0.04$ & $76^{25}_{25}$ & $1.03^{0.40}_{0.40}$ & -- \\
A1767 & 0.0701 & $4.1^{1.8}_{0.9}$ & 7 & $3.83^{3.53}_{2.51}$ & $-24.84 \pm {0.39}$ & $3.96 \pm 0.07$ & $59^{45}_{27}$ & $1.27^{1.00}_{0.64}$ & -- \\
A2107 & 0.0421 & $4.31^{0.57}_{0.35}$ & 22 & $3.61^{1.53}_{1.39}$ & $-25.04 \pm {0.28}$ & $4.53 \pm 0.06$ & $56^{18}_{16}$ & $1.35^{0.52}_{0.47}$ & -- \\
A2147 & 0.0351 & $4.91^{0.18}_{0.18}$ & 42 & $5.69^{1.84}_{1.84}$ & $-24.62 \pm {0.19}$ & $5.19 \pm 0.03$ & $60^{15}_{15}$ & $1.27^{0.42}_{0.42}$ & -- \\
A2151 & 0.0370 & $2.4^{0.06}_{0.06}$ & 23 & $5.32^{1.86}_{1.86}$ & $-25.87 \pm {0.29}$ & $5.18 \pm 0.14$ & $19^{4}_{4}$ & $3.65^{1.18}_{1.18}$ & -- \\
A2440 & 0.0904 & $3.88^{0.16}_{0.14}$ & 7 & $3.11^{1.55}_{1.55}$ & $-25.79 \pm {0.34}$ & $6.15 \pm 0.15$ & $35^{9}_{9}$ & $2.14^{0.72}_{0.72}$ & -- \\
A2589 & 0.0416 & $3.7^{1.38}_{0.69}$ & 19 & $6.74^{4.99}_{3.34}$ & $-24.29 \pm {0.29}$ & $3.39 \pm 0.03$ & $59^{38}_{23}$ & $1.27^{0.86}_{0.56}$ & -- \\
A2593 & 0.0433 & $3.1^{1.5}_{0.9}$ & 23 & $7.17^{6.44}_{4.37}$ & $-24.91 \pm {0.28}$ & $4.56 \pm 0.05$ & $33^{26}_{17}$ & $2.22^{1.84}_{1.25}$ & -- \\
A2626 & 0.0573 & $2.9^{2.5}_{1}$ & 14 & $7.58^{11.56}_{5.46}$ & $-24.95 \pm {0.34}$ & $4.70 \pm 0.07$ & $29^{40}_{17}$ & $2.52^{3.54}_{1. 62}$ & -- \\
A2634 & 0.0312 & $3.7^{0.18}_{0.18}$ & 27 & $5.18^{1.83}_{1.83}$ & $-24.53 \pm {0.26}$ & $3.33 \pm 0.03$ & $60^{16}_{16}$ & $1.24^{0.41}_{0.41}$ & -- \\
A3389 & 0.0265 & $2.1^{0.9}_{0.6}$ & 27 & $8.27^{6.67}_{4.92}$ & $-25.05 \pm {0.31}$ & $3.29 \pm 0.07$ & $24^{18}_{12}$ & $2.82^{2.11}_{1.56}$ & -- \\
\tableline
\vspace{-9 mm}
\tablecomments{Columes: (1) Name; (2) Redshift; (3) Emission--weighted
  mean temperature; (4) Estimated number of member galaxies; (5) Characteristic
  number density; (6) Characteristic magnitude; (7) Total luminosity within 
  $r_{500}$; (8) $K$--band mass--to--light ratio; (9) Stellar mass fraction; 
  (10) ICM mass fraction. 
  Uncertainties in all columns quoted at $1\sigma$ level. $M_*$ \& $\phi_*$ 
  calculated assuming $\alpha = -1.1$. All calculations done with $h_{70}=1$.
  }
\end{tabular}
}
\end{center}
\vskip-20pt
\end{table*}

\begin{inlinefigure}
   \ifthenelse{\equal{\figtype}{EPS}}{
   \begin{center}
   \epsfxsize=8.cm
   \begin{minipage}{\epsfxsize}\epsffile{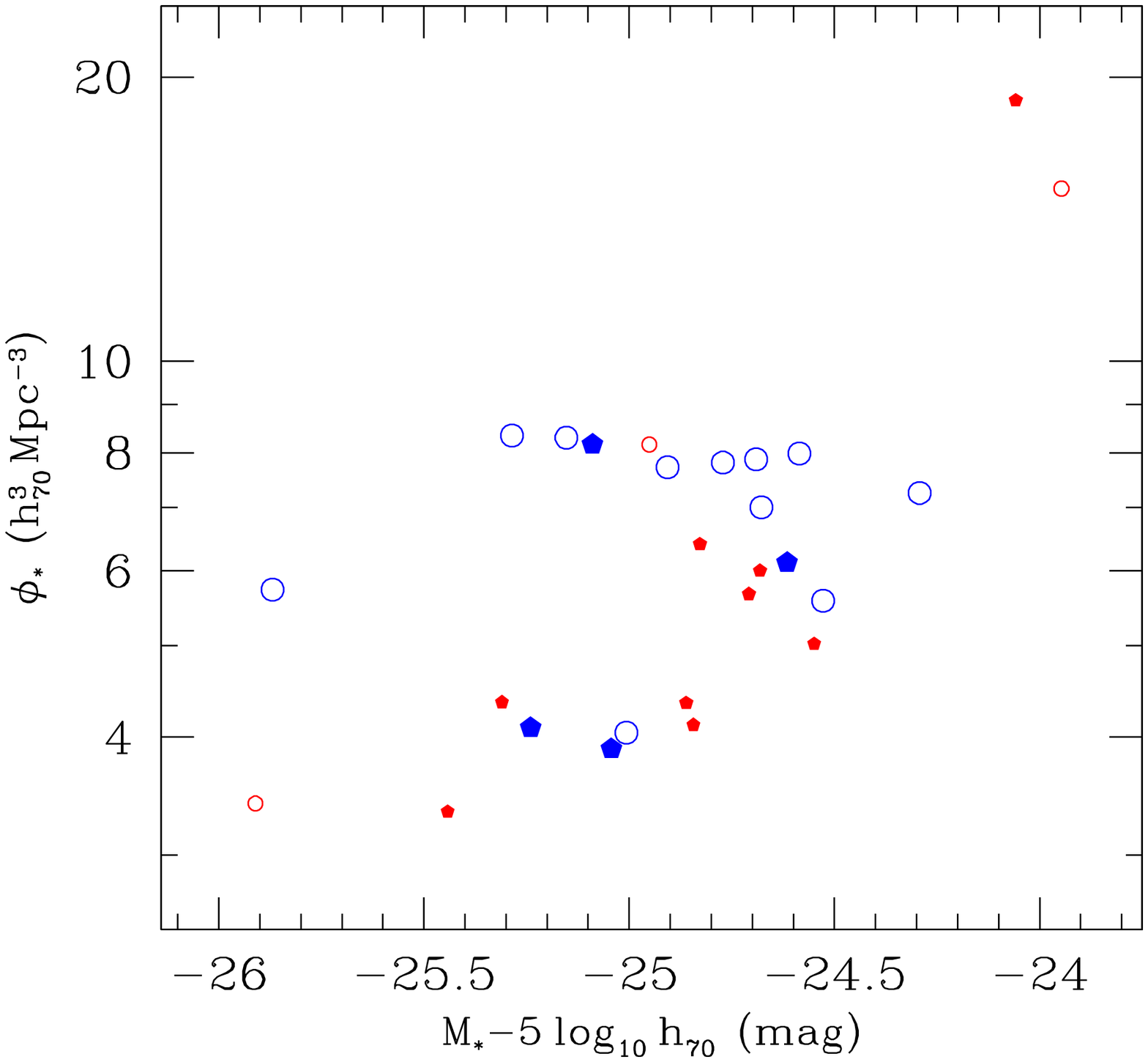}\end{minipage}
   \end{center}}
   {\myputfigure{f2.pdf}{0.0}{1.0}{-70}{-40}}
   \figcaption{\label{fig:lstar} Measured $M_*$ and $\phi_*$ for our cluster 
     sample (see Table 1). Solid circles are clusters with $kT \ge 4$\,keV.
     Larger symbols denote clusters with $z \le0.05$, while smaller
     symbols refer to $z>0.05$ clusters.  }
\end{inlinefigure}

\subsubsection{Luminosity Function}
\label{sec:lf}

We transform the apparent magnitudes into solar luminosities using 
$K_{\odot} = 3.39$ \citep{kochanek01}, the cluster distance and Galactic 
extinction: 
\begin{equation}
M_{K} = K -25-5\, {\rm log_{10}}(d_L/1{\rm Mpc})-A_{K},
\end{equation}
where $d_L$ is the luminosity distance, and $A_{K}$ accounts for the
Galactic extinction. We ignore k--corrections, because 
our clusters are all at very low redshifts.  The Galactic extinction, which has a small effect on the overall luminosity, is assumed to be the same for the entire cluster; we obtain the 
extinction values from the NASA/IPAC Extragalactic Database .

Using the NFW model with 
concentration $c=5$, we deproject to convert the projected number and luminosity into 
an estimate of the number 
$N_{obs}$ and luminosity $L_{obs}$ of galaxies within $r_{500}$.  The 
deprojection essentially reduces the background corrected luminosity and number 
by  $20-30$\%.  These quantities divided by the cluster volume $V_{cls}=(4\pi/3)
r_{500}^3$ are the cluster mean number density $n_{obs}$ and luminosity density 
$j_{obs}$, which are related to the cluster luminosity function $\phi(L)$:
\begin{equation}
\label{eq:schechter}
\nonumber
n_{obs}  = \int_{L_{lim}}^\infty \phi(L)\, dL \ \ {\rm and}\ \ 
j_{obs}  = \int_{L_{lim}}^\infty L\, \phi(L)\, dL,
\end{equation}
where $L_{lim}$ is the luminosity corresponding to $K_{lim}$ for
each particular cluster. We use the \citet{schechter76} form
for the luminosity function:
\begin{equation}
\phi(L)\,dL = \phi_* \left( {L \over L_*} \right)^\alpha\, {\rm e}^{-L/L_*} d\left({L \over L_*}\right),
\end{equation}
where $L_*$ and $\phi_*$ are characteristic luminosity and number densities, 
respectively, and $\alpha$ is the faint-end power law index. 

In practice, the 2MASS data do not go deep enough to provide useful constraints 
on the faint end slope $\alpha$ for all of our clusters.  Thus, we use external
constraints to set $\alpha$.  The \nir cluster Schechter function
faint end slope is generally found to be $-1.3 \lesssim \alpha \lesssim -0.8$
\citep[e.g.][]{mobasher98,andreon00,tustin01,balogh01a,depropris02}. In the following we 
adopt $\alpha = -1.1$. (We discuss the effect of different $\alpha$ on 
our results in \S\ref{sec:err}.)  Given $n_{obs}$, $j_{obs}$, $\alpha$ and 
$L_{lim}$,  we solve for $\phi_*$ and $L_*$ using Eqn~\ref{eq:schechter}.

Fig~\ref{fig:lstar} contains the derived $\phi_*$ and $M_*$
(the absolute magnitude corresponding to $L_*$).  The distribution of 
$M_*$ ($\overline{M_*}-5\log h_{70}=-24.88\pm 0.02$) in Fig~\ref{fig:lstar} and Table 1 shows no clear correlation 
with cluster mass or redshifts, but many nearby clusters exhibit larger $\phi_*$.   
Our cluster sample is based on existing \xray cluster catalogs, and so
there is a tendency for the low temperature systems to lie at low redshift. For example, the mean redshift of the lower temperature half of the sample (i.e. $kT_X <4$~keV)  
is $0.040\pm0.001$ as compared to $0.060\pm0.001$ for the other half.  
This could potentially pose a problem, because any redshift related systematic could masquerade as a mass related trend in our cluster population.  We probe for systematics by examining the luminosity function parameters within the hot and cold  cluster subsamples.  Dividing each subsample at redshift 0.05, we find no statistically significant differences with redshift in $\overline{M_*}$ or $\overline{\phi_*}$ for either of the subsamples.  However, within the same redshift range, $\overline{\phi_*}$ is significantly higher for the low temperature subsample, consistent with the shallow light--mass relation presented above.

In obtaining the Schechter parameters by using the observed light and number of galaxies, we do not consider the brightest cluster galaxy (BCG), which typically contains a significant fraction of the cluster light and would bias the parameters of the cluster luminosity function.  We account for the BCG light separately, modeling cluster luminosity functions with Schechter functions plus the light of the BCG.
There are two clusters whose Schechter parameters appear to be different
from that of other clusters: A133 \& A1651. Their $M_*$ are relatively dim, and
their $\phi_*$ about twice as large as others. This may be because of their
relatively small number of member galaxies, and their extraordinarily bright
central dominant galaxies: removal of the luminosity contribution from the BCGs
dramatically reduces $L_*$ and boosts $\phi_*$. These two clusters underscore the fact that our integral approach to solving for  $L_*$ and $\phi_*$ leads to correlated errors in $L_*$ and $\phi_*$ that maintain an accurate estimate of the total cluster light even when estimates of  $L_*$ and $\phi_*$ individually can be quite inaccurate.

\subsubsection{Estimating Total Light and Stellar Mass}
\label{sec:totlite}

With the Schechter parameters $\phi_*$ and $L_*$, we estimate the
total cluster luminosity by integrating over the luminosity function: 
$L_{tot} = V_{cls}\, L_*\, \phi_*\,\Gamma(\alpha+2) + L_{bcg}$. To reduce 
the dependence of our light estimate on the faint end slope parameter $\alpha$ 
we truncate the luminosity function at the absolute magnitude $M_{low}= -20$:
\begin{equation}
\label{eq:jtot}
j_{tot}'  =  L_{bcg}/V_{cls} + \int_{L_{low}}^\infty L\, \phi(L)\, dL.
\end{equation}
For our sample and choice of $\alpha$, $j_{tot}'$ differs from $j_{tot}$ only at $2-3\%$ level. In the
analyses that follow we make no distinction between these two, and
use the ``truncated'' luminosity as our luminosity estimate throughout.

We also apply a correction for the fact that the the default 2MASS
magnitudes underestimate the total light of the galaxies. By comparing
the 2MASS default (isophotal) magnitudes of their galaxy sample with
the total magnitudes of deep photometry \citep{pahre99},
\citet{kochanek01} found a systematic difference of 20\%.  
We multiply $L_{tot}$ by the same factor of 1.2.

To obtain the total stellar mass, we multiply the total luminosity $L_{tot}$ by 
the average stellar \mlr $\overline{\Upsilon}_{star}(T_X)$ for each cluster.
We estimate the mean stellar \mlr from the observations of elliptical
and spiral galaxies \citep{gerhard01,bell01}, and take into account 
the varying spiral fraction as a function of cluster temperature 
\citep[][notice that the spiral fractions were determined using $B$-band light 
rather than $K$-band; we neglect possible wavelength dependence of the spiral
fraction here]{bahcall77b,dressler80b}.
The details of our approach are presented in the
Appendix.  We show in Fig~\ref{fig:mlstar} the resulting mean stellar
\mlr $\overline{\Upsilon}_{star}$, and compare it with the result 
from the 2dF galaxy redshift survey \citep{cole01}, which was obtained by using 
the observed galaxy colors and redshifts to constrain their
evolution, based on the recent version of a population synthesis code 
\citep{bruzual93}. 
This value  $0.73 \Upsilon_\odot$ (where $\Upsilon_\odot$ is the solar $K$-band
\mlre) is consistent with our lowest temperature clusters 
(highest spiral fraction), but over the full 
range of cluster masses examined in this sample, the stellar \mlr 
is expected to vary by $\sim$10\%.

\begin{inlinefigure}
   \ifthenelse{\equal{\figtype}{EPS}}{
   \begin{center}
   \epsfxsize=8.cm
   \begin{minipage}{\epsfxsize}\epsffile{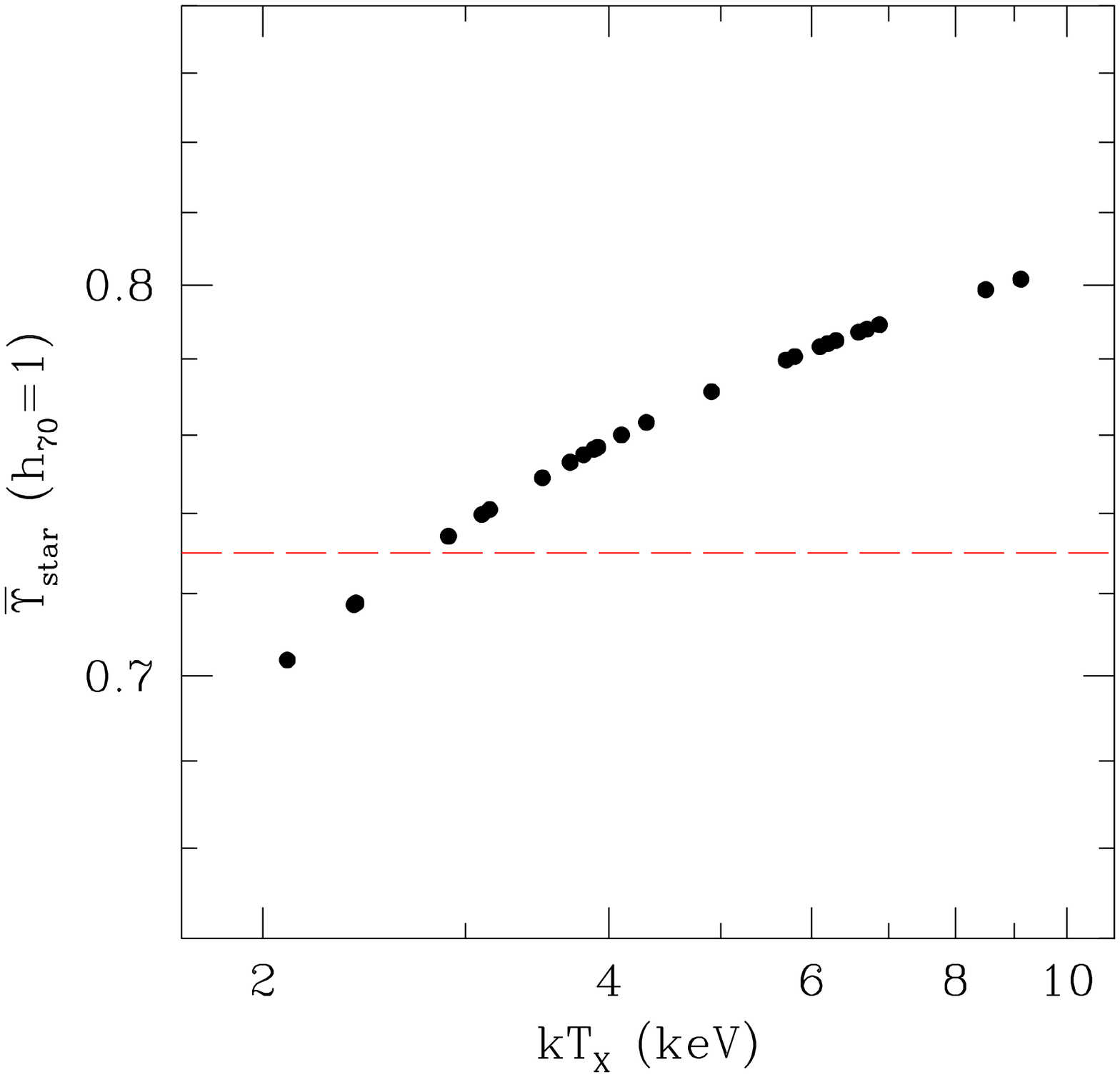}\end{minipage}
   \end{center}}
   {\myputfigure{f3.pdf}{0.0}{1.0}{-90}{-30}}
   \figcaption{\label{fig:mlstar} 
   The $K$-band mean stellar \mlr $\overline{\Upsilon}_{star}(T_X)$
   based on the observed \mlr in elliptical and
   spiral galaxies and spiral fractions in clusters. See Appendix
   for more details. The dashed-line is the mean value from the
   2dF survey \citep{cole01} $\overline{\Upsilon}_{star,2dF} = 0.73
   \,\Upsilon_\odot$, assuming a \citet{kennicutt83} intial mass function.
     }
\end{inlinefigure}

\subsection{Estimating the ICM Mass from X--ray Emission}
\label{sec:geticm}

We calculate the ICM mass for the 13 clusters in the MME subsample
using fits to the ROSAT Position Sensitive Proportional Counter (PSPC)
\xray surface brightness fits from the literature
\citep{mohr99}.  The 0.5:2.0~keV \xray emissivity is relatively
insensitive to temperatures (varying by $10\%$ for a factor of two
increase in temperature for clusters with temperatures above 1.5~keV),
so one needs not know the cluster temperature profile to determine the
ICM density profile from an \xray image.  We have adjusted the density
profiles to reflect improved temperature measurements in two of the
clusters.  Analysis of mock PSPC images of simulated galaxy clusters
indicates that cluster ICM masses can be estimated with an accuracy of
$\sim$10\% \citep{mohr99} and that clumping in the ICM enhances the
\xray emissivity leading to an estimated $\sim10$\% bias in ICM mass
estimates \citep{mohr99,mathiesen99}.  We correct for this bias by
reducing ICM mass estimates for all 13 clusters by 10\%.  However,
this reduction is counterbalanced when one considers the $\sim 10\%$
diminution of the cluster baryon fraction at $r_{500}$ compared to the
global baryon fraction $\Omega_b/\Omega_M$ as seen in numerical
studies, where $\Omega_b$ is the total baryon
density in the universe.  We recalculate the ICM masses for this
analysis using the  $M_{500}-T_X$ relation adopted for the rest of the analysis
\citep[][see Eqn~\ref{eq:m-t}]{finoguenov01}.

\section{NIR Luminosity and Cosmology}
\label{sec:nirlite}

One critical issue for ongoing \nir cluster surveys is determining
the degree to which the \nir light from the stellar populations
within cluster galaxies traces the galaxy cluster binding mass.
Cluster binding mass or virial mass is central to using the abundance
of galaxy clusters to constrain cosmology \citep[although deep surveys
contain enough information to solve for cluster binding mass and
cosmology simultaneously;][]{majumdar02}.  If \nir galaxy light is
a good indicator of cluster binding mass, then large solid angle
surveys like 2MASS could be used to (1) define very large samples of
clusters in the nearby universe \citep[see][]{kochanek02}, and (2)
provide direct constraints on a combination of the local power
spectrum of density fluctuations and the mean matter density.  Due to
the limited size of our sample, we simply focus on the relation
between the virial mass (inferred from \xray observations) and stellar
luminosity (obtained from 2MASS photometry) of nearby clusters,
discussing the implications for future \nir cluster surveys.  In
addition, we use the \nir light and measurements of cluster ICM masses
to address the baryon fraction, the cluster \mlre, and the
cosmological matter density parameter.

\begin{inlinefigure}
   \ifthenelse{\equal{\figtype}{EPS}}{
   \begin{center}
   \epsfxsize=8.cm
   \begin{minipage}{\epsfxsize}\epsffile{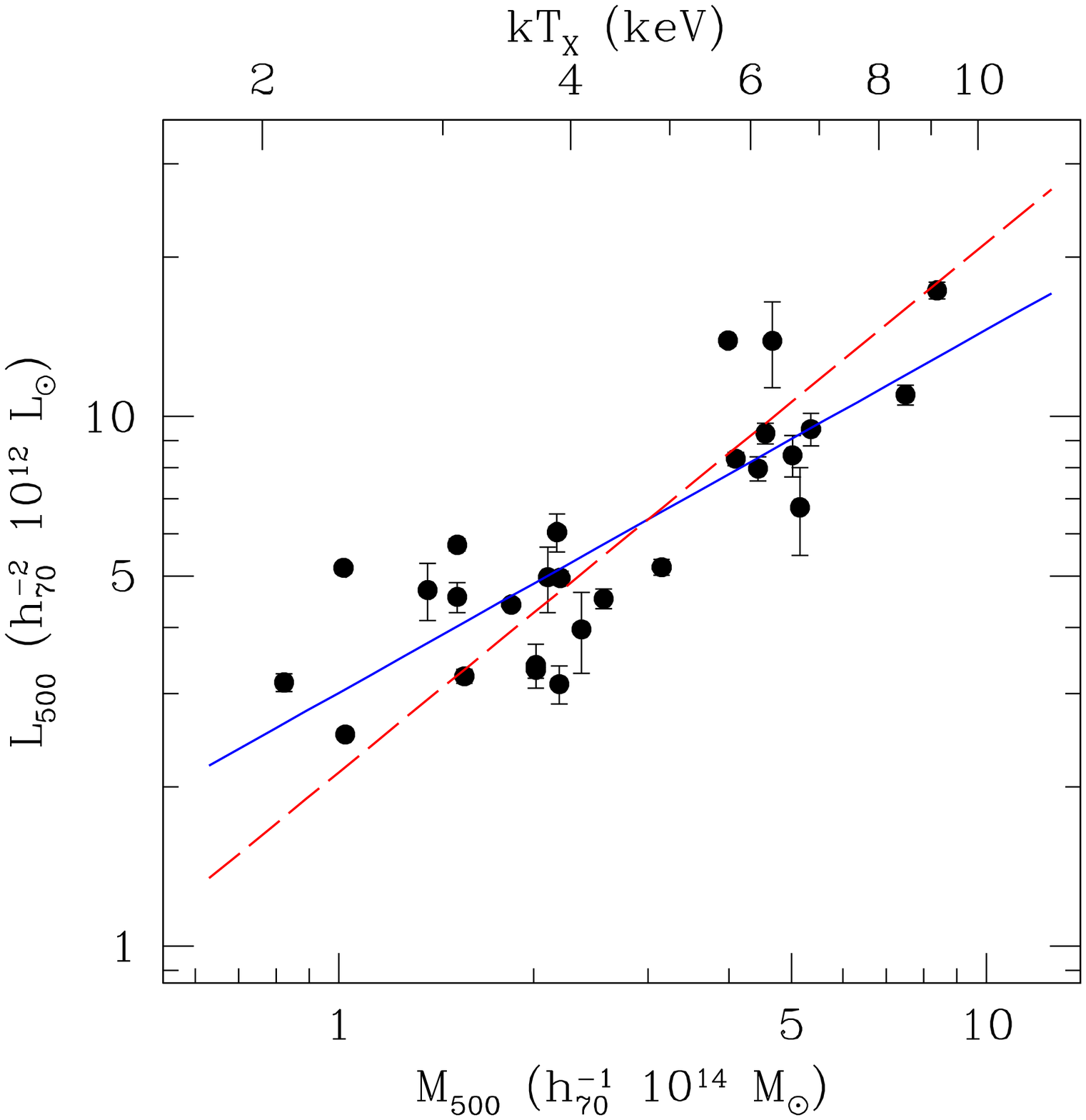}\end{minipage}
   \end{center}}
   {\myputfigure{f4.pdf}{0.0}{1.0}{-70}{-20}}
   \figcaption{\label{fig:LT}
        The cluster $K$-band luminosity--mass correlation. The uncertainties
	in $L_{500}$ are rather small, typically less than 10\% (see Table 
	1). The solid line shows the best-fit relation of slope of $0.69$ 
       (Eqn~\ref{eq:lm}), while the dashed line shows a slope of $1.0$ behavior.
           }
\end{inlinefigure}

Figure~\ref{fig:LT} contains a plot of \nir light $L_{500}$ versus the
binding mass $M_{500}$, corresponding to the light and mass contained
within the cluster radius $r_{500}$.  We use Eqn~\ref{eq:MT} to
estimate $M_{500}$, with the cluster emission-weighted mean
temperature $T_X$ taken from the literature.  The corresponding
temperature scale is presented at the top of the figure. For our
sample of 27 nearby clusters there is a reasonably tight correlation
between cluster mass and $K$-band light
\begin{equation}
\label{eq:lm}
L_{500}=6.4\pm 0.4\,\times 10^{12}\,{L_\odot \over h_{70}^2}
    \left({M_{500}\over 3\times 10^{14}\,h_{70}^{-1}\,M_\odot}\right)^{0.69\pm 
0.09}.
\end{equation}
The best--fit results in this paper are obtained by minimizing the 
vertical distances in log space to the line from the data points, and no uncertainties in 
$M_{500}$ are included in the fitting procedure. The 1$\sigma$ uncertainties are
determined by bootstrap resampling and refitting $10^3$ times. As implied by the
equation, the pivot point for the fit is at $3\times 10^{14}\,h_{70}^{-1}\,M_\odot$, roughly
corresponding to $4.8$~keV.

The best-fit correlation (solid line) differs from
$L_{500}\propto M_{500}$ (dashed line) by $\sim3\sigma$, indicating a \mlr that
increases with mass.  In essence, low mass clusters produce relatively
more light per unit binding mass than high mass clusters.  Possible
explanations include a higher star formation efficiency in low mass
clusters (see $\S$\ref{sec:starformation}) or a stripping
process that liberates stars from galaxies more efficiently in high
mass than in low mass clusters \citep[e.g.][]{trentham98,gregg98}.  As 
discussed in $\S$\ref{sec:intro},
we do not expect the $K$-band stellar \mlr to be very sensitive to
stellar population and star formation history variations, but gross 
differences in the stellar populations of low and high mass clusters 
could in principle contribute to the shallow \nir light--mass 
relation found here.  We return to this issue in $\S\ref{sec:icmstar}$.

The fractional scatter in $L_{500}$ about this best fit relation is
28\%, suggesting that $K$-band light does indeed provide a useful
tracer of cluster mass.  Possible sources of the scatter include: the
variations in the cluster star formation and galaxy formation history, variations in the 
faint-end slope of the luminosity function $\alpha$, the scatter ($\sim 17\%$) in the adopted 
$M-T_X$ relation, the projection of unusually bright background galaxies onto the clusters, and the
``deprojection'' of the cluster luminosity (i.e. deviations from spherical symmetry and from the assumed galaxy distribution model).
In practice, the uncertainty in binding mass estimates will depend on the
filter used to detect and characterize clusters.  
Crudely speaking, using the slope and amplitude of the scaling relation,
for a given $K$-band light, one could predict the galaxy cluster mass with
a statistical accuracy of $\sim45$\%.
Of course this is likely a lower limit to the true
mass uncertainties in a \nir cluster survey, because in practice one will
not know the apparent virial radius $\theta_{500}$ and cluster position \emph{a priori}.

Figure~\ref{fig:M2L} contains the $K$-band \mlr versus binding mass $M_{500}$.  
The trend of increasing $K$-band \mlr $\Upsilon_{500}$ is best fitted by
\begin{equation}
\label{eq:MoverL}
{\Upsilon_{500}} = 47 \pm 3\,h_{70}\,\Upsilon_\odot
    \left({M_{500}\over 3\times 10^{14}\,h_{70}^{-1}\,M_\odot}\right)^{0.31\pm 0.09}.
\end{equation}
The Spearman correlation coefficient for the fit is 0.63, with a 
$1.6 \times10^{-3}$ probability of being consistent with no correlation.
The average \mlr over all clusters is $47\pm 3\,h_{70}\,\Upsilon_\odot$ 
(short dashed line) and the average for clusters with $kT_X \ge 3.7$~keV (more 
massive than 
$2\times 10^{14}\,h_{70}^{-1}\,M_\odot$) is $53\pm 3\,h_{70}\,\Upsilon_\odot$.
A fit to the clusters with $kT_X\ge3.7$~keV produces a slope that is 
consistent with zero ($0.03 \pm 0.12$), which is one way of demonstrating that 
much of the leverage on the slope comes from the low mass systems.
The 8 clusters that have $kT_X < 3.7$~keV are of low redshift ($0.016 \le z
\le 0.057$, $\overline{z} =0.034$).  Because visual inspections of these cluster
fields do not show unusual behavior relative to higher $T_X$ clusters (e.g. 
deficiency of galaxies), we believe the low 
$\Upsilon_{500}$ for these low $T_X$ clusters is robust.

In general, trends in blue/visual band mass--to--light ratio with temperature 
have not been seen \citep[e.g.][]{hradecky00,david95}, but more recently 
both \citet{bahcall02} and \citet{girardi02} reported positive correlations
between cluster blue band ($V$ and $B$, respectively) \mlr and
temperature. \citet{bahcall02} observed $\Upsilon_V \propto
T_X^{0.3\pm0.1}$, which is somewhat shallower than the trend we see in
our sample ($\Upsilon_K \propto T_X^{0.5 \pm 0.1}$); the
best-fit of \citet{girardi02} sample is $\Upsilon_B \propto M^{0.25}$.
Our results are statistically consistent with these other relations. 
\citet{bahcall02} interpret the dependence on temperature as
an aging effect in the stellar population.  If that is the case, then one
would expect this trend to be less prominent in the \nire, where for
old stellar populations the stellar \mlr is relatively insensitive to
population age.  Because we see the trend in the $K$-band, it is
certainly possible that there is an underlying variation in star
formation efficiency over cluster mass scales (we examine this possibility
in $\S\ref{sec:starformation}$).

\begin{inlinefigure}
   \ifthenelse{\equal{\figtype}{EPS}}{
   \begin{center}
   \epsfxsize=8.cm
   \begin{minipage}{\epsfxsize}\epsffile{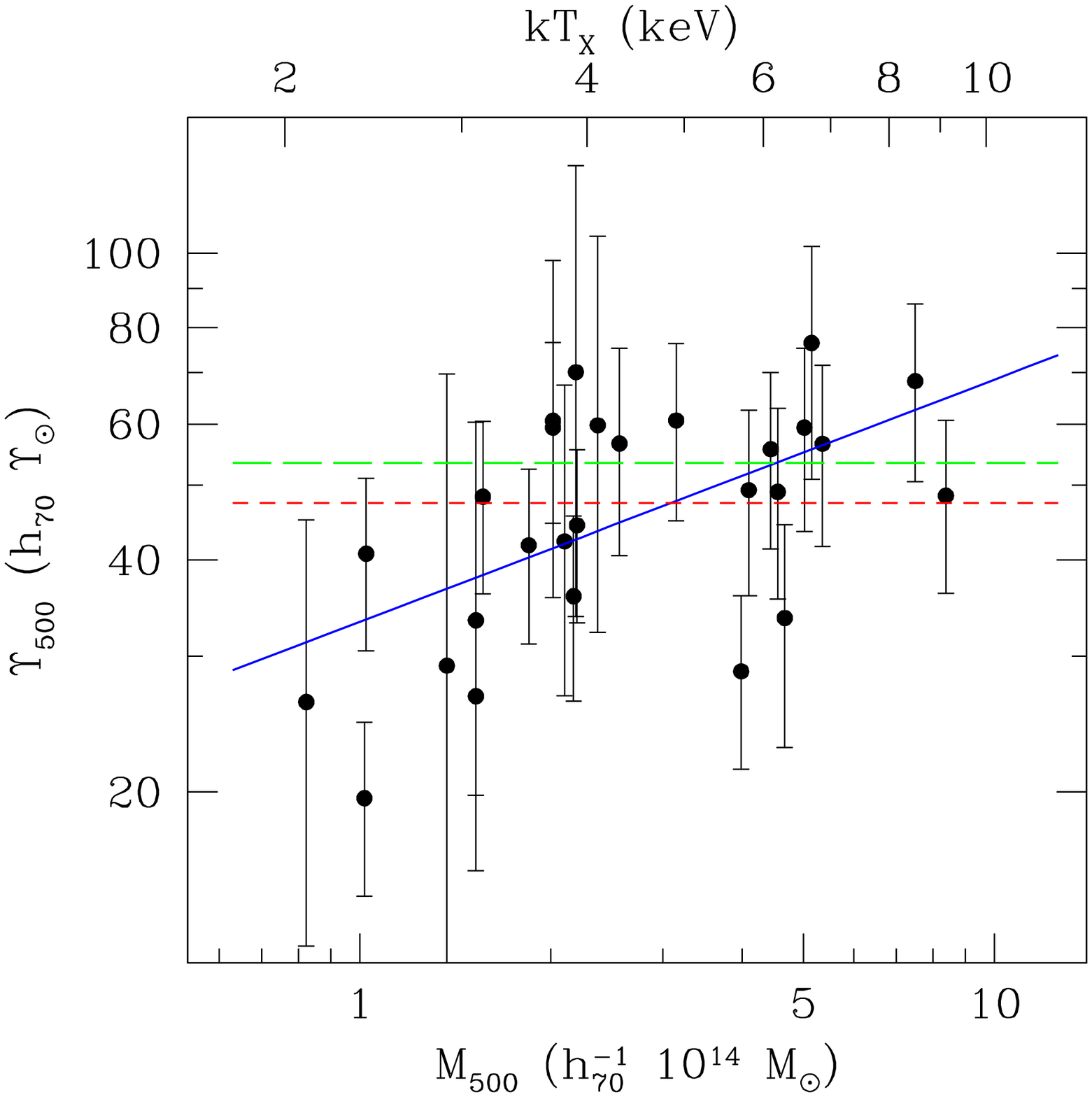}\end{minipage}
   \end{center}}
   {\myputfigure{f5.pdf}{0.0}{1.0}{-70}{-20}}
   \figcaption{\label{fig:M2L}
      The cluster $K$-band \mlr vs cluster mass. The solid line is the best-fit
      relation Eqn~\ref{eq:MoverL}, the short-dashed line is the average
      of all clusters -- $47 \pm 3\, h_{70} \Upsilon_\odot$, and the long-dashed
      line is the average of hot ($kT_X \ge 3.7$ keV, or $M_{500} \gtrsim 
      2\times \,10^{14} h_{70}^{-1} M_\odot$) clusters -- $53 \pm 3\, h_{70} 
      \Upsilon_\odot$.
      The error bars mainly reflect the uncertainties in cluster binding
      mass estimated from $T_X$.
           }
\end{inlinefigure}

Our results are consistent with the $K$-band \mlr found in a detailed
spectroscopic and 2MASS photometric study of the Coma cluster ($T_X = 8.21
\pm0.16$ keV):  $\Upsilon = (68\pm 21)\,h_{70}\, \Upsilon_\odot$ \cite[][when
converted from their projected \mlr to the ``deprojected'' \mlr adopted in our 
analysis]{rines01b}.  Interestingly, \citet{kochanek02} find a weakly decreasing
\mlr using 2MASS data.  Evaluating cluster $K$-band light at $r_{200}$, they 
find $\Upsilon_{200} \propto M_{200}^{-0.10\pm0.09}$, which is inconsistent
with our results at about the $4\sigma$ level.  Their approach relies on a 
pseud--matched filter cluster finding algorithm using an
NFW model with fixed virial radius and constant $L_*$ and $\alpha$.  The number of galaxies within the cluster virial region $N_{*666}$ is estimated by rescaling the number observed using a fixed virial radius NFW filter.  They show that $N_{*666}$ is correlated with several estimators of binding mass.  Although both our approach and theirs should, in principle, give 
consistent results, we believe our method offers several
advantages for studies of the $K$-band light -- virial mass relationship. First, we start with a proven, low scatter \xray estimator of cluster virial mass, and we use \xray emission peaks as the cluster center.  Thus, we know the size and location of our cluster virial region, whereas in their approach they have to pull these quantities from the 2MASS data and bootstrap to an estimate of the total cluster light.  Second, we use the NFW model only in deprojecting the cluster light, whereas their cluster detection algorithm is built upon the NFW model.  Thus, our method would be less affected by deviations in some clusters from the NFW model.

\subsection{Cluster Mass-to-Light Ratio Constraint on $\Omega_M$} 
\label{sec:M2L}

Cluster scale \mlrs $\Upsilon_{500}$, together with
observations of the mean luminosity density of the universe
$\overline{j}$, have often been used to estimate the mean matter
density $\overline{\rho}$ or matter density parameter $\Omega_M$ 
\citep[e.g.][among others]{bahcall77,trimble87,bahcall95,carlberg96,
rines01b,bahcall02,girardi02}. Taking the mean \mlr of the universe to be
$\Upsilon_{univ}=\overline{\rho}/\overline{j}$, then
$\Omega_M \approx\overline{j}\Upsilon_{500}/\rho_c$, if the cluster \mlr
is equal to the universal \mlre.  A potential weakness of this approach
is that the galaxy populations inside and outside galaxy clusters
differ significantly, with field galaxies being predominantly late
type with a mix of young and old stellar populations and cluster
galaxies being predominantly early type with mostly old stellar
populations \citep[e.g.][]{dressler84}.  With these
dramatic galaxy population differences, it is not clear that the star
formation efficiency within clusters would be representative of that
in the universe as a whole.  Nevertheless, an application of this
technique using $K$-band light is particularly interesting, because it
minimizes \mlr differences due to the ages of stellar populations.  Perhaps 
most interesting is that in combination with external constraints on the matter 
density parameter, these results will provide insights into the star formation 
history in clusters as compared to the universe as a whole (see 
\S\ref{sec:systematics}).

Because the $K$-band cluster \mlr is an increasing function of 
cluster mass, we will evaluate $\Omega_M$ by using
both the mean $\overline{\Upsilon}_{500,all} = (47\pm3)\,h_{70}\,\Upsilon_\odot$
of all clusters, and the mean \mlr of $T_X \ge 3.7$~keV clusters, 
$\overline{\Upsilon}_{500,hot} = (53\pm 3)\,h_{70}\, \Upsilon_\odot$.
For the mean luminosity density, we adopt the value obtained by
\citet{kochanek01}:
$\overline{j}=(5.00 \pm 0.52)\times 10^8\,h_{70}\,L_\odot$ Mpc$^{-3}$.
Combining $\overline{\Upsilon}_{500,all}$ or $\overline{\Upsilon}_{500,hot}$ 
with $\overline{j}$, with the critical density $\rho_c = 1.36\times 10^{11}
\,h_{70}^2\,M_\odot\,$Mpc$^{-3}$, we estimate the matter density parameter 
$\Omega_M=0.17 \pm 0.02$ (whole sample) or $\Omega_M=0.19\pm 0.03$ (hot 
cluster subsample).

\subsection{Total Baryon Fraction Constraint on $\Omega_M$}
\label{sec:totbaryon}

Another way of constraining the mean matter density of the universe
from the clusters is by using the baryon fraction in galaxy clusters
to estimate the universal baryon fraction.  In combination with
external constraints on the baryon density parameter, one can then
estimate the matter density parameter
\citep[e.g.][MME]{swhite93a,david95,evrard97}.
Hydrodynamical simulations indicate that the ICM is more extended in
its distribution than the cluster dark matter, and so corrections for
this ``depletion'' effect must be included.  In addition, hydro
simulations indicate that clumping and substructure in the ICM enhance
the \xray emission relative to that expected if the gas were smoothly
distributed \citep[MME;][]{mathiesen99}.  A correction for this
``clumping'' effect must also be included.

Here we only consider the two dominant baryon reservoirs in galaxy
clusters: stars in the galaxies and the hot ICM (that is, we do not
specifically include a contribution of the interstellar medium in
galaxies, intergalactic stars or dark baryonic objects). We write the
cluster baryon fraction as $f_b \equiv (M_{star}+M_{ICM})/M_{500}$, where
$M_{star}$ and $M_{ICM}$ are the total stellar and ICM mass inside $r_{500}$, 
respectively, and we have corrected the ICM contribution for depletion and 
clumping (see \S\ref{sec:geticm}).  We estimate $M_{star}$ using the $K$-band 
luminosity (see \S\ref{sec:totlite}).  If the corrected cluster baryon fraction 
then reflects the global one, we have $\Omega_M = \Omega_b/f_b$. 

\begin{inlinefigure}
   \ifthenelse{\equal{\figtype}{EPS}}{
   \begin{center}
   \epsfxsize=8.cm
   \begin{minipage}{\epsfxsize}\epsffile{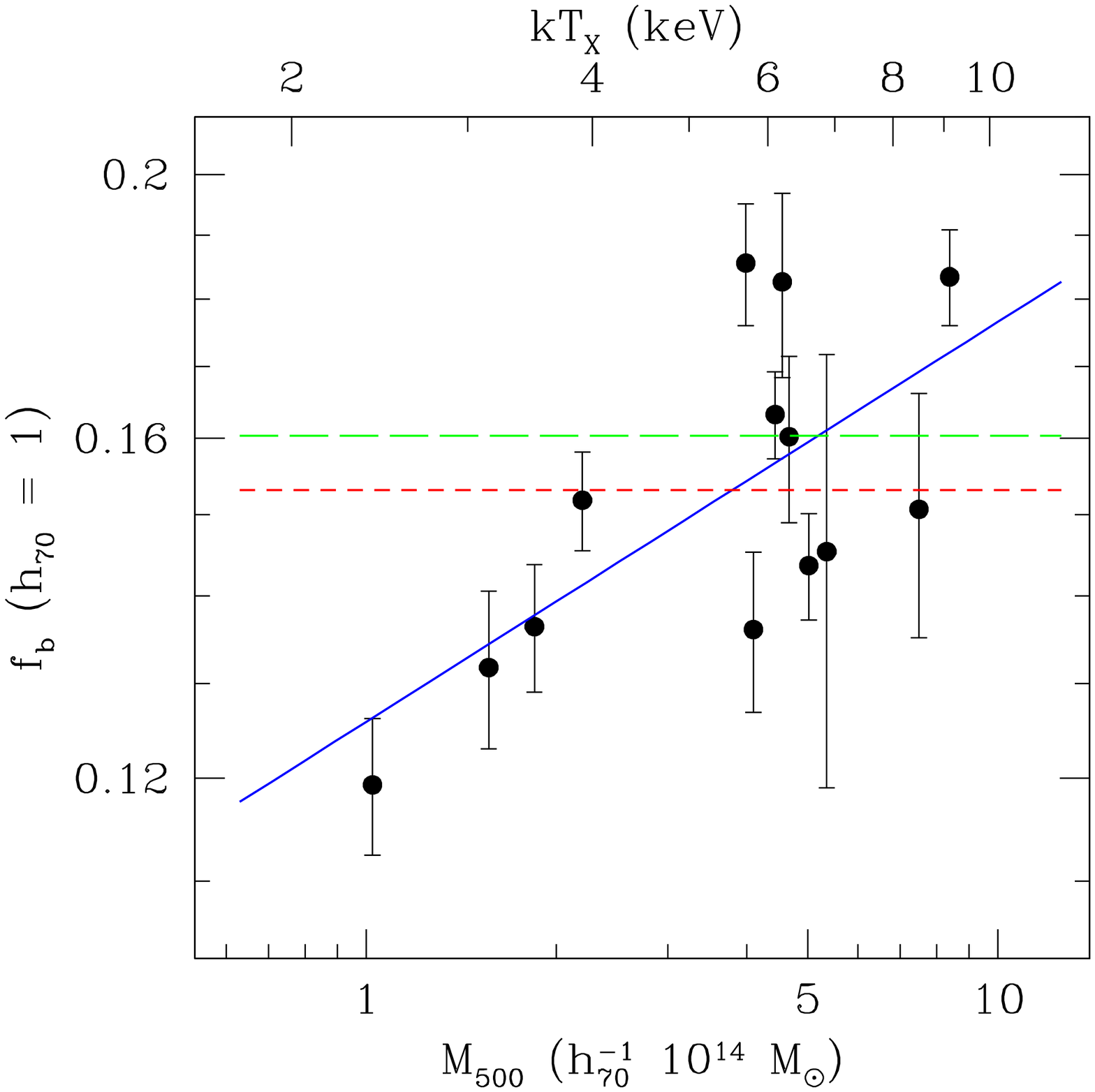}\end{minipage}
   \end{center}}
   {\myputfigure{f6.pdf}{0.0}{1.0}{-70}{-20}}
   \figcaption{\label{fig:bsum}
      The baryon fraction $f_b$ as a function of cluster mass for the 13 
      2MASS clusters in the MME subsample. The long-dashed line is the mean of 
      the 10 clusters that have $kT_X \ge 3.7$\,keV, $\overline{f}_{b,hot} = 
      0.160$. The short-dashed line is the mean for all 13 clusters, 
      $\overline{f}_{b,all} = 0.153$.
           }
\end{inlinefigure}

We show in Fig~\ref{fig:bsum} the total baryon fraction in our cluster sample,
which is an increasing function of cluster mass.  The best-fit is
\begin{equation}
f_b = 0.148_{-0.004}^{+0.005}
  \left({M_{500}\over 3\times 10^{14}\,M_\odot}\right)^{0.148\pm 0.04}
\end{equation}
for $h_{70}=1$.
Note that this sample includes only 13 clusters -- namely, those clusters 
in our ensemble that also lie in the MME \xray flux limited sample of 45 
clusters.  The increase in the baryon fraction with cluster mass mainly reflects
the dependence of the ICM mass fraction on cluster mass, because the ICM mass 
dominates over the stellar mass in galaxy clusters (we examine this issue in 
\S\ref{sec:star2icm} below).
This correlation between ICM mass and temperature was noted by MME, whose
study favors the trend $M_{ICM}/M_{500}
\propto T_X^{0.34\pm 0.22}$, which is a likely indication of preheating 
\citep[e.g.][]{bialek01} or perhaps variations in star formation efficiency 
\citep{bryan00b}. Because our depletion correction from hydrodynamical 
simulations is most appropriate for massive clusters, and because massive 
clusters are the systems that sample the largest portions of the universe, 
we take the baryon fraction in our massive systems as most representative 
of the universal baryon fraction. In fact, the best-fit slope for the 
$f_b - T_X$ relation of the hot ($kT_X \ge 3.7$~keV) clusters in this subsample 
is much shallower and consistent with zero: $0.054 \pm 0.152$, which 
may be an indication of the cluster baryon fraction asymptotically 
approaches the universal baryon fraction.
Following the approach of the 
last section, we will present constraints on the density parameter that 
arise from the average characteristics of the entire subsample and the hot 
clusters in this subsample (10 clusters).

The mean baryon fraction for the 10 clusters with $kT_X \ge 3.7$~keV
is $\overline{f}_{b,hot} = 0.160 \pm 0.006$, and for the whole MME
subsample we have $\overline{f}_{b,all} = 0.153 \pm 0.006$.  We do not present 
the $h$ scaling, because $M_{star}$ and $M_{ICM}$ scale with the Hubble constant
differently.  With the value of $\Omega_b\,h_{70}^2=0.0457\pm0.0018$ from 
{\it WMAP} measurements, we estimate $\Omega_M$; for the hot
clusters in the subsample we get $\Omega_M = 0.28\pm 0.03$, while
for the whole MME subsample $\Omega_M = 0.30\pm 0.03$.  This value
is consistent with the previous results using the same technique.
\citet{evrard97} deduced that $\Omega_M < 0.40$ using the previous
\xray observations. Based on ICM mass fraction measurements of 45 \& 30 clusters, MME and
    \citet{ettori99} concluded that $\Omega_M < 0.32$ \& $0.34$, respectively.
\citet{grego01} estimated $\Omega_M \sim
0.25$ from the SZE
observations of 18 clusters. Finally, with Chandra observations of 7
intermediate redshift clusters, \citet{allen02} deduced that $\Omega_M
= 0.30^{+0.04}_{-0.03}$.

\begin{table*}[htb]
\begin{center}
\caption{Derived $\Omega_M$}
\begin{tabular}{ccc}
\tableline \tableline
Method & Sample & $\Omega_M$ \\
\tableline
mass--to--light & total (27) & $0.17\pm 0.02$ \\
ratio $\Upsilon_{500}$ & massive (19) & $0.19\pm 0.03$\\
\tableline
baryon & MME (13) & $0.30\pm 0.03$ \\
fraction $f_b$ & MME, massive (10) & $0.28 \pm 0.03$ \\
\tableline
\end{tabular}
\tablecomments{Numbers in parentheses indicate the number of clusters in each 
   sample/subsample. ``massive'' denotes clusters more massive than 
   $2\times 10^{14}\,M_\odot$. All quantities calculated assuming $h_{70}=1$. }
\end{center}
\vskip-20pt
\end{table*}

\subsection{Comparison of $\Omega_M$ Constraints}
\label{sec:systematics}

Our baryon fraction estimate of $\Omega_M$ is in excellent agreement with {\it WMAP} 
constraints from the CMB anisotropy: $\Omega_M(f_b) = 0.28 \pm 0.03$ versus
 $\Omega_M(WMAP) = 0.27 \pm 0.04$ \citep{bennett03}.  Because the analysis that leads to the mass--temperature relation we use to estimate our cluster virial masses is so different from the X-ray imaging analysis required to estimate the ICM masses, the agreement between our baryon fraction estimate of $\Omega_M$ and the {\it WMAP} estimate suggests that our cluster virial masses and ICM masses must be fairly accurate.
In that case we can use the baryon fraction \Om estimate to constrain differences between the universal 
and the cluster \mlre.  Specifically, with the universal \mlr 
$\Upsilon_{univ} = \Omega_M \rho_c
/\overline{j}$ where $\Omega_M$ is given by cluster baryon fraction, we have
$\Upsilon_{univ} = (78 \pm 13)\,h_{70}$, whereas our hot cluster \mlr is 
$\overline{\Upsilon}_{500,hot} = (53 \pm 3)\,h_{70}$. Expressed slightly
differently, our measurements imply that 
$\Upsilon_{univ}/\overline{\Upsilon}_{500,hot}=1.46\pm 0.21$. 
It is well known that galaxy morphological mix and star formation rates
are very different between the fields and the clusters \citep[e.g.][]
{dressler84,lewis02}; our measurements thus provide an 
interesting indication of possible differences in the star formation histories 
and star formation efficiencies in these two dramatically different 
environments.

\section{Star Formation and Enrichment}
\label{sec:icmstar}

Combined analyses of \nir and \xray observations allow us to address
several interesting questions concerning the thermodynamic history of baryons in the clusters:
the star formation efficiency (\S\ref{sec:starformation}), the mass
fractions of stars and ICM (\S\ref{sec:star2icm}), and the metal
enrichment of the ICM (\S\ref{sec:enrich}).  We discuss the implications
of our results on it in \S\ref{sec:thermo}.

\subsection{Variation of the Star Formation Efficiency}
\label{sec:starformation}

Understanding the cosmic star formation history is of fundamental
importance in understanding the formation and evolution of galaxies
\citep[e.g.][]{madau98,springel03}. Here we estimate the total stellar
mass fraction for our sample of 27 clusters.  This quantity can be
regarded as a tracer of the total mass in stars formed in galaxies
within the cluster halo over the course of its collapse.  The total
stellar mass for each cluster is converted from the total $K$-band
luminosity, using the derived mean stellar \mlr (see 
\S\ref{sec:totlite} \& Appendix). Because \nir light is relatively
insensitive to the star formation history, our data should provide a
robust estimate of the total stellar mass (within galaxies) in the clusters.

\begin{inlinefigure}
   \ifthenelse{\equal{\figtype}{EPS}}{
   \begin{center}
   \epsfxsize=8.cm
   \begin{minipage}{\epsfxsize}\epsffile{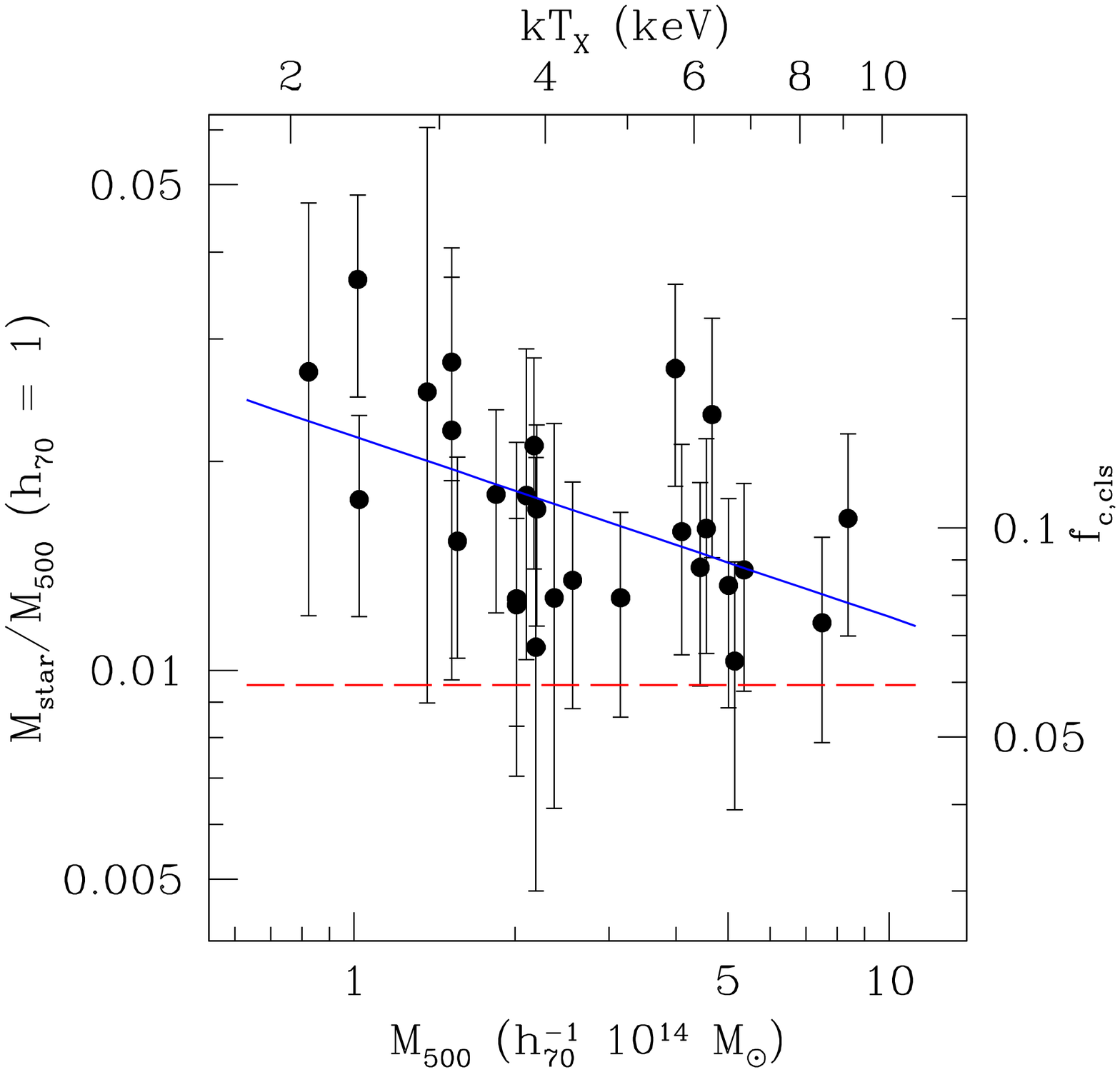}\end{minipage}
   \end{center}}
   {\myputfigure{f7.pdf}{0.0}{1.0}{-70}{-20}}
   \figcaption{\label{fig:starform}
        Fraction of halo mass that has been turned into stars by the present 
	epoch in our cluster sample. 
	The error budgets are dominated by the uncertainties in the cluster
	binding mass, which are due to the uncertainties in the \xray 
	temperature measurement and in the $M-T_X$ relation (see Table 1).
	On the right axis is shown the cooled
	baryon fraction in the clusters (see \S\ref{sec:star2icm} for 
        definition). The dashed line indicates the universal cold baryon 
	fraction (see \S\ref{sec:star2icm}).
           }
\end{inlinefigure}

Figure~\ref{fig:starform} is a plot of the fraction of the halo
mass which is in stars within galaxies for each of our clusters.  This
fraction varies from $\sim 2.2\%$ at low mass scales of
10$^{14}h_{70}^{-1}\,M_\odot$ to $\sim 1.2\%$ at mass scales of
10$^{15}h_{70}^{-1}\,M_\odot$.  It is important to note that this
fraction does not include any stars that have been stripped from
their parent galaxies and are freely orbiting within the galaxy
cluster potential, as has been observed in a some cases \citep{ferguson98}. 
Also shown in the figure is the best-fit relation
\begin{equation}
{M_{star} \over M_{500}} = 1.64_{-0.09}^{+0.10} \times 10^{-2}
  \left({M_{500}\over 3 \times 10^{14}\,M_\odot}\right)^{-0.26\pm 0.09}
\end{equation}
for $h_{70}=1$.  Our observations suggest that the integrated star formation 
efficiency drops by a factor of $\sim1.8$ over the mass range of galaxy 
clusters.   Of course, it is also possible that the trend we observe arises 
because the reservoir of stars that lie outside galaxies is fractionally larger 
in more massive clusters \citep[e.g.][]{trentham98,gregg98}.

The decrease in star formation efficiency appears to be broadly consistent with theoretical 
expectations from state-of-the-art hydrodynamical simulations by
\citet{springel03}. Their model includes radiative cooling and heating of
the gas, a multi-phase description of the interstellar medium, 
a self-regulating star formation mechanism, and the feedback processes from
supernova events and galactic outflows. They study star formation within haloes
ranging from $10^8$--$10^{15} M_\odot$, and find star formation
rates broadly consists with observations. 
In their models, the integrated star formation efficiency as a function of halo mass falls 
by a factor of 5 to 10 over cluster mass scales.  The underlying cause is the less efficient formation of cooling flows in halos with virial temperatures above $10^7$\,K
\citep{springel03}. However, what we observe is the star formation efficiency over all galaxy--mass
halos that lie within our cluster halo at the present epoch.  Because
at early times the star formation efficiency in galaxies will be
comparable for those galaxies in low mass and high mass cluster halos,
we expect the theoretical prediction for $M_{star}/M_{500}$ to suggest
a much weaker trend.  We are currently making a detailed comparison of
our observational results with numerical star
formation models.

\subsection{Variation of the ICM to Stellar Mass Ratio}
\label{sec:star2icm}

Another interesting quantity that has received much attention is the
ICM to stellar mass ratio, $M_{ICM}/M_{star}$, as a function of
cluster mass. This quantity can be recast as the cluster cold baryon fraction,
$f_{c,cls} = M_{star}/(M_{star}+M_{ICM}) = (1+M_{ICM}/M_{star})^{-1}$, which is
relevant in (numerical) studies of cosmic star formation 
\citep[see][and references therein]{balogh01b}.
In this section we examine these quantities for the MME subsample.

Our joint analysis of \nir and \xray observations shows 
that $M_{ICM}/M_{star}$ is an increasing function of cluster mass, varying
from a factor of 5.9 for low mass clusters to a factor of 10.4 for
high mass clusters. The best-fit relation for $h_{70}=1$ is
\begin{equation}
{M_{ICM} \over M_{star}} = 7.7_{-0.4}^{+0.5}
    \left({M_{500} \over 3\times 10^{14}\,M_\odot} \right)^{0.25\pm 0.07}
\label{eq:icm2star}
\end{equation}
Over an order of magnitude in mass, there is roughly a two fold increase
in $M_{ICM}/M_{star}$; the existence of a trend in these data is
significant at the 3.5\,$\sigma$ level, providing confirmation of
the early work by \citet{david90}, who estimated the more massive
clusters ($T_X \sim 7\,$keV) have a $\sim 3 \times$ larger
$M_{ICM}/M_{star}$ ratio than do less massive ones ($T_X \sim
3\,$keV). \citet{arnaud92} found a correlation between $M_{ICM}$ and
stellar luminosities from early-type galaxies (E$+$S0); they estimated
the mean ICM to stellar mass ratio as $3.6$, which is much smaller
than the values that we obtain. We note these two early studies
evaluated the ICM to stellar mass at different overdensities for different 
clusters.
More recently, \citet{roussel00} found
no trend when they examined $M_{ICM}/M_{star}$ at $r_{200}$ in their combined
\xray and optical sample of 33 clusters and groups.  Interestingly,
their analysis suggested a trend when measurements were made at
$r_{2000}$, where less extrapolation of the \xray data is required.

\begin{inlinefigure}
   \ifthenelse{\equal{\figtype}{EPS}}{
   \begin{center}
   \epsfxsize=8.cm
   \begin{minipage}{\epsfxsize}\epsffile{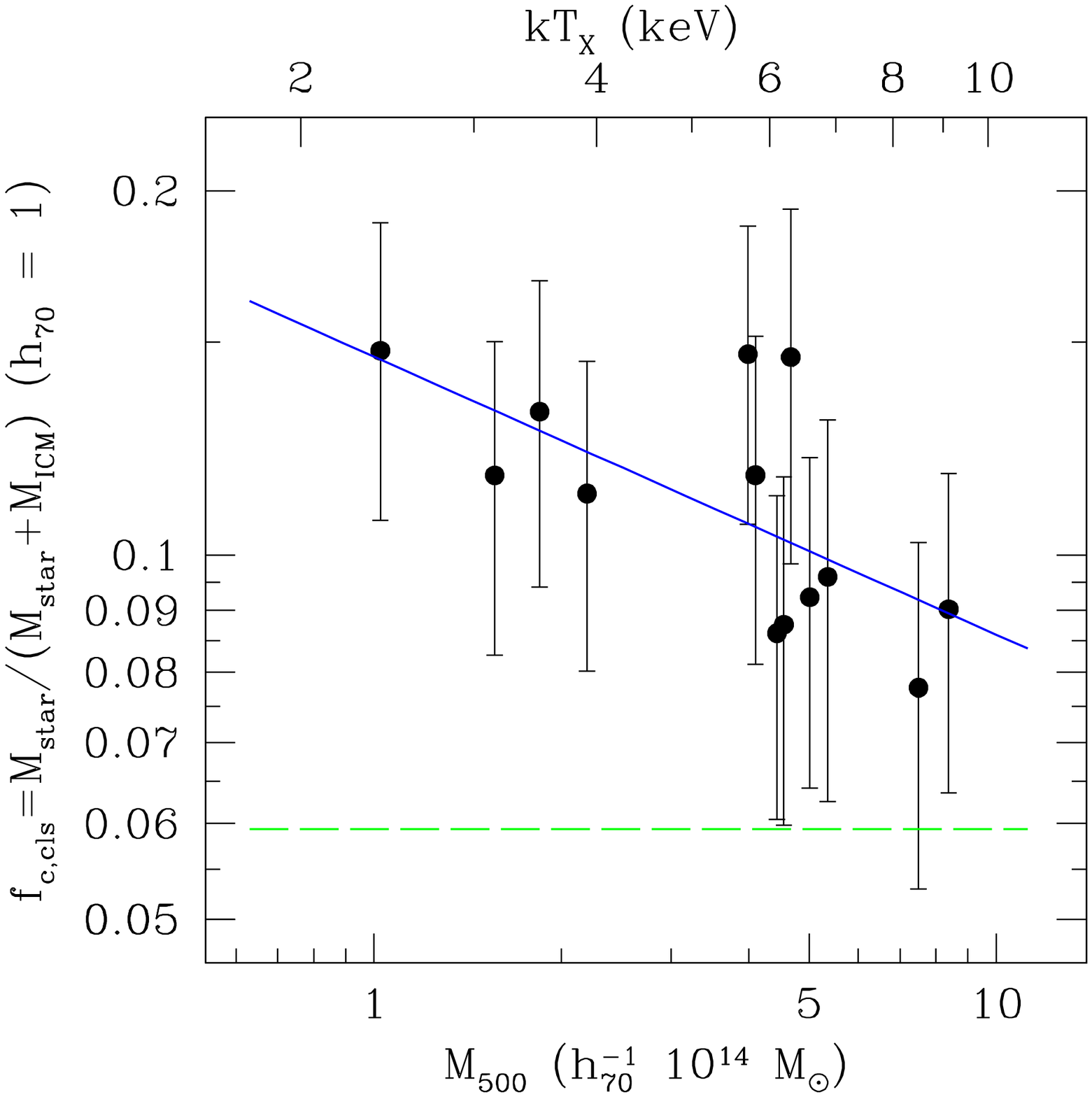}\end{minipage}
   \end{center}}
   {\myputfigure{f8.pdf}{0.0}{1.0}{-70}{-20}}
   \figcaption{\label{fig:f2f}
        The cold baryon fraction of the clusters. This is roughly inversely
        proportional to the ICM to stellar mass ratio. The solid line is the 
	best-fit to the observed trend. The dashed line shows the universal
	cold baryon fraction.
           }
\end{inlinefigure}

In Fig~\ref{fig:f2f} we plot the cluster cold baryon fraction, the ratio
of stellar mass to total baryon (star plus ICM) mass. As implied by 
Eqn~\ref{eq:icm2star}, $f_{c,cls}$ is a decreasing function of cluster mass,
indicating that the star formation efficiency is smaller in more massive 
clusters than that in low mass ones. This is in accordance with the conclusion
of \S\ref{sec:starformation}; in fact, from the ratio $M_{star}/M_{500}$ one
can infer the cold baryon fraction via $f_{c,cls} = (M_{star}/M_{500})\,
\Omega_M/\Omega_b$ \citep{balogh01b}, which is shown in the right axis of
Fig~\ref{fig:starform}. Also shown in both Figures~\ref{fig:starform} \& 
\ref{fig:f2f} (the horizontal dashed line) is the universal cold fraction 
$f_{c,univ} = \Omega_{star}/\Omega_b$, where $\Omega_{star} = (2.7\pm 0.3)\times
10^{-3}\,h_{70}^{-1}$ is obtained by an analysis of 2MASS data 
\citep{kochanek01}. We notice that in both figures the cluster cold fraction 
approaches the universal value with increasing halo mass.

Our Figures~\ref{fig:starform} \& \ref{fig:f2f} can be directly compared to the
bottom and upper panel of Fig 1 in \citet{balogh01b}, which was produced using 
previous observations of 48 clusters and groups \citep{carlberg96,roussel00}. 
In that study, it was argued that the first approach (which utilizes the ratio
$M_{ICM}/M_{star}$) produces large scatter for $f_{c,cls}$, while the second
one (that uses the ratio $M_{star}/M_{500}$) shows a much smaller scatter, and
implies cluster cold fractions for all clusters considered are close to the
universal value. However, measurements in our sample show this may not be the case.
Our results may be more robust,
because of the data homogeneity (all NIR photometry based on 2MASS), an improved treatment of stellar \mlre, the use of observed \xray $M-T$ relation (as opposed
to that based on numerical simulations), and the reliability of the ICM mass 
estimation (all \xray data extend to $r_{500}$).  Studies of a larger sample will provide additional insights into these trends.

\subsection{ICM Enrichment}
\label{sec:enrich}

Extensive investigations of ICM metal abundances have been carried out (see
\citealt{arnaud92, renzini97} for reviews), but until recently these studies
relied on emission--weighted values and were therefore biased toward the 
abundance at the cluster center.  Precision abundance measurement and radial 
abundance profiles in
clusters have only been possible with imaging \xray observatories such as ASCA,
BeppoSax, Chandra and XMM-Newton \citep[e.g.][among others]{dupke00,
finoguenov00, finoguenov01b,degrandi01}.

In the bottom panel of Fig~\ref{fig:IMLR}, we plot the iron yield, which is the ratio of iron ($Fe$) to stellar mass, as a function of cluster mass in units of solar metalliticity:
$y_{cls,Fe} \equiv M_{Fe}/M_{star} = (M_{ICM}\,Z_{ICM,Fe}+M_{star}\,Z_{star,Fe})/M_{star}$
\citep[e.g.][]{arnaud92}.
The ICM iron mass is calculated by assuming the mean mass weighted iron
abundance of the ICM to be $Z_{ICM,Fe} = 0.21 Z_{\odot,Fe}$, which
is deduced from the metallicity profiles in cool core clusters and
non--cool core clusters observed by BeppoSax 
\citep[][2003 private communication]{degrandi01}. 
For the stellar iron abundance, we account for population trends in the 
galaxies much as we do in estimating the typical stellar \mlre.  We assume  
constant metallicities $Z_{star,Fe}
=1.2 Z_{\odot,Fe}$ and $1.8 Z_{\odot,Fe}$ for ellipticals and spirals 
\citep{jorgensen99,terlevich02}, respectively, and we weight them
by the appropriate stellar \mlre, luminosity functions of different 
morphological types, and the relative abundance of ellipticals and spirals. We 
detail our approach in the Appendix.  The resultant $Z_{star,Fe}$ is about 
$1.45 Z_{\odot,Fe}$ at $T_X =2$ keV and about $1.27 Z_{\odot,Fe}$ at $T_X=10$ 
keV.

The best-fit to the observed trend in our data is
\begin{equation}
y_{cls,Fe} = 2.98_{-0.09}^{+0.10}\,Z_{\odot,Fe}\left( {M_{500} \over 3\times 10^{14}\,M_\odot}\right)^{0.12 \pm 0.04}
\end{equation}
for $h_{70}=1$. 
The iron yield is related to the iron \mlr (IMLR, 
\citealt{renzini93}) which appears frequently in recent literature (e.g. 
\citealt{finoguenov00}): IMLR $= y_{cls,Fe} \overline{\Upsilon}_{star}$,
where $\overline{\Upsilon}_{star}$ is the mean stellar \mlre.
Previous studies using blue band optical photometry suggested that IMLR is generally independent of cluster mass \citep{renzini97}.  The trend in our sample for increasing $y_{cls,Fe}$ arises from the increasing ICM mass fraction,  an improved understanding of the typical ICM metallicity \citep{degrandi01} and the decreasing stellar mass fraction.  
Fig~\ref{fig:IMLR} (lower panel) also shows that the iron 
yields in the cluster systems are very high relative to the approximately 
solar metal abundances observed in stars and the interstellar medium in galaxies
\citep[and the references therein]{henry99}.
It is unlikely that galaxies with a simple star formation history and
typical initial mass function (hereafter IMF) could  produce such high iron yields 
(although see \citealt{pipino02}).

\begin{inlinefigure}
   \ifthenelse{\equal{\figtype}{EPS}}{
   \begin{center}
   \epsfxsize=8.cm
   \begin{minipage}{\epsfxsize}\epsffile{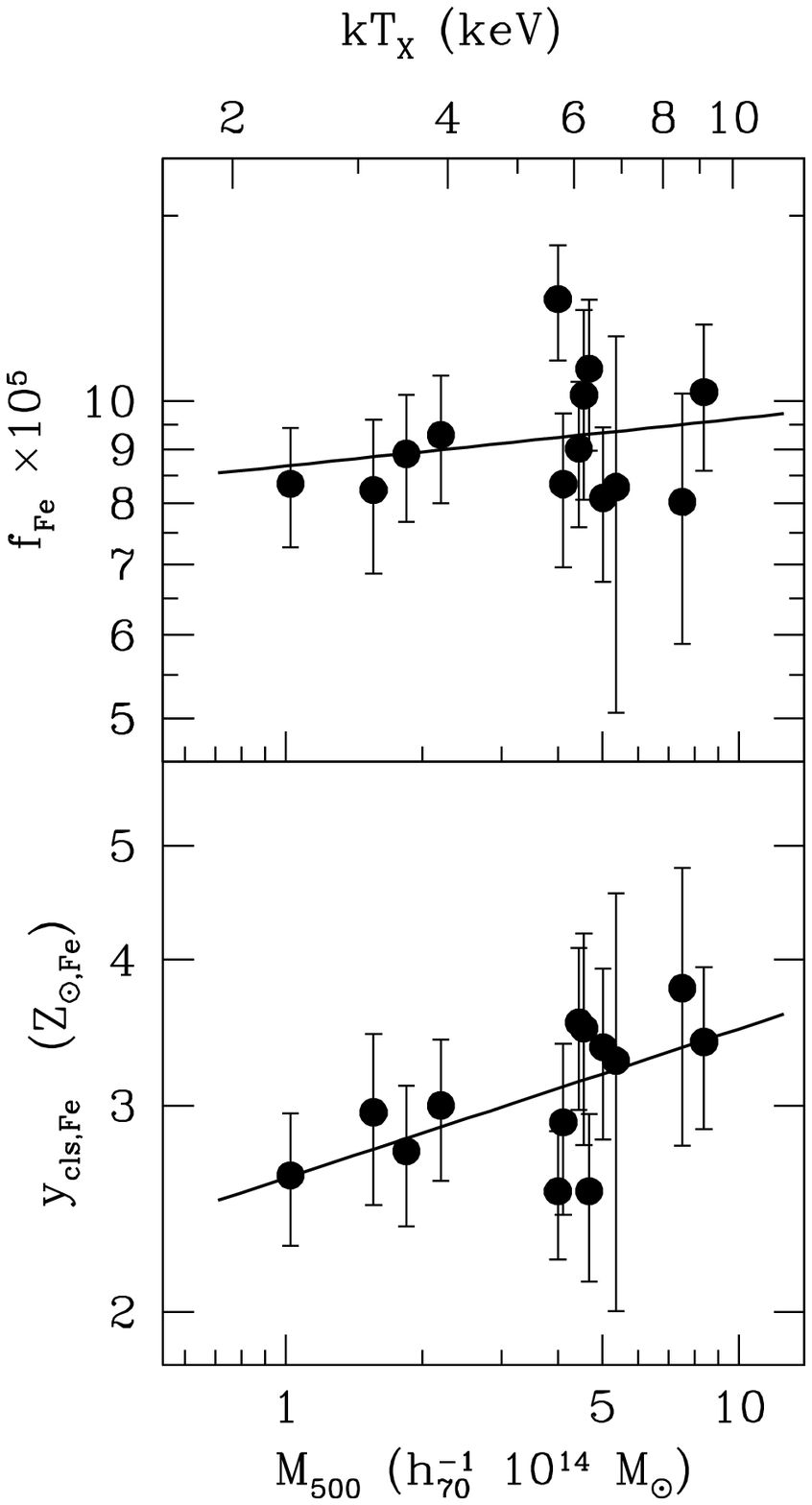}\end{minipage}
   \end{center}}
   {\myputfigure{f9.pdf}{1.0}{0.7}{120}{-500}}
   \figcaption{\label{fig:IMLR}
	The upper panel: the iron mass fraction in the clusters.  The lower 
	panel: the iron yield.  The ICM iron abundance 
	$Z_{ICM,Fe} = 0.21\,Z_{\odot,Fe}$ from \citet{degrandi01} is used, and 
	the stellar contribution is calculated by the method stated in the 
	Appendix.
        }
\end{inlinefigure}

In Fig~\ref{fig:IMLR} (upper panel) we plot the iron mass
fraction, $f_{Fe} = M_{Fe}/M_{500}$, using $Z_{\odot,Fe} = 1.814\times 10^{-3}$
\citep{anders89}. Unlike the iron yield, which compares the iron mass to the 
stellar mass in each system, the iron mass fraction shows less apparent trend 
with respect to the cluster binding mass;
the best--fit slope of the iron mass fraction is $0.048\pm 0.056$,
consistent with no trend at all. 
This nearly uniform enrichment of the baryon reservoir
in clusters, combined with the high iron yield and the trend of decreasing
star formation efficiency with cluster mass, may suggest that the primary
enrichment mechanism in the ICM is not the stars that we observe in
the cluster galaxies; instead, the intergalactic gas that becomes the 
ICM may be pre--enriched by some earlier
stellar population that is at work before cluster galaxies form.
This process appears to work with similar efficiency over
the whole range of cluster masses. One possible 
mechanism for this uniform enrichment would be population III stars.  An early 
stellar population of massive stars could contribute significantly to the 
metallicity -- both because their likely top-heavy IMF and high metal yields 
\citep[e.g.][]{larson98,heger02} -- and presumably enhance the entropy of the 
gas that eventually collapses to form galaxies and clusters. 
Another intriguing piece of cluster evidence for population III stars is that 
the observed elemental abundance 
ratios in the cluster ICM appear to be difficult to explain by reasonable 
combinations of Type I and Type II supernovae \citep{loewenstein01}.

\begin{table*}[htb]
\begin{center}
\caption{Scaling Relations}
\begin{tabular}{ccc}
\tableline \tableline
Quantity & Normalization         & Slope \\
X        & $a$ &  $b$\\
\tableline
$L_{500}$          &$3.0^{+0.4}_{-0.3} 10^{12} L_\odot$ &$0.63\pm 0.09$ \\
$\Upsilon_{500}$   & $32\pm 4\,\Upsilon_\odot$          &$0.31\pm 0.09$ \\
$f_b$              & $0.13 \pm 0.08 $                   &$0.15\pm 0.04$ \\
$f_{star}$         & $0.022^{+0.003}_{-0.002}$          &$-0.26\pm 0.09$\\
$M_{ICM}/M_{star}$ & $5.9^{+0.8}_{-0.7}$                &$0.30\pm 0.07$ \\
$y_{Fe}$           & $2.6\pm 0.2\,Z_{\odot,Fe}$         &$0.12\pm 0.04$ \\
\tableline
\end{tabular}
\tablecomments{Scaling relations in the form of $X = 
  a\,(M_{500}/10^{14}\,M_\odot)^b$. Uncertainties at $1\sigma$ level.
  All quantities calculated assuming $h_{70}=1$.}
\end{center}
\vskip-20pt
\end{table*}

\subsection{Thermodynamic History of the ICM}
\label{sec:thermo}

Clearly the intersection of cluster structure and the cosmic star formation 
history is an intriguing avenue for further research.
The star formation efficiency, the ratio of ICM mass to stellar mass and the 
total iron mass provide important insights into the thermodynamic history of the
ICM.  These results, taken together with existing observations of other cluster 
scaling relations (e.g. $L_X - T_X$, $M_{ICM} - T_X$, $R_I-T_X$, where $R_I$ is
the \xray isophotal size of clusters) provide an abundance of evidence that cluster ICM and stellar properties do not scale self--similarly 
\citep[e.g. MME,][]{david93,mohr97a,arnaud99,lloyd-davies00}.  One
interpretation has been that the gas distribution in low mass clusters
has been altered relative to the expectations of the standard
formation scenario by the addition of entropy or ``preheating'' by
galaxy formation at early times \citep{kaiser91,evrard91,cavaliere97,wu00,bialek01,muanwong02}.
In this picture, prior to the cluster/group collapse, the intergalactic gas is
heated by, e.g., supernova--driven galactic outflows or AGN.  The
preheated, high--entropy gas thus forms a more extended distribution
when it collapses into the cluster scale halos. The effect is less
prominent in more massive systems, because in these massive systems 
the entropy increase in the ICM
due to strong accretion shocks is much larger than that caused by
preheating.  This framework provides a means of explaining the observed slopes of the 
 $M_{ICM} -T_X$, $L_{X} - T_X$ and $R_{I} - T_X$ relations, if preheating enhances  the entropy by about $\sim$100~keV\,cm$^2$ \citep{bialek01}.  This level of
preheating appears to be consistent with the cluster ``entropy floor''
observed by \citet{lloyd-davies00}.

Despite the successes of the preheating model in explaining observed
cluster ICM properties, a generic problem with this model is the 
large energy required if carried out in the dense
ICM in the centers of clusters or in the early universe
\citep[e.g.][]{bryan00b,wu00}.  To circumvent these
difficulties, \citet{bryan00b} proposed an alternative to the
preheating model in which the star formation efficiency varies as a
function of halo mass.  The gas that cools to form stars is the lowest
entropy gas, and its removal from the cluster causes similar
structural signatures to those produced in the preheating scenario
\citep[see also][]{voit02,muanwong02,nath02b}.

The \nir and \xray properties of our cluster ensemble appear to
provide problems for both the preheating and star formation efficiency
scenarios.  Namely, the direct evidence for a decreasing star
formation efficiency with increasing cluster mass
(Fig~\ref{fig:starform}) lies outside the bounds of simple preheating
scenarios, and the increasing total baryon fraction is at odds with
simple models where the total baryon fraction is constant in clusters
but the fraction cooling into stars is changing.  Moreover, the high
metallicity of the ICM together with the large ICM mass to stellar
mass ratio strongly suggests that the source of enrichment (and so
perhaps the entropy as well) requires something more than star
formation with a typical initial mass function.  It appears likely that
more complex models that include both varying star formation
efficiency, preheating and enrichment at early times are necessary to
explain these observations.

\section{Systematic Uncertainties}
\label{sec:err}

There are several key ingredients of our analysis: the cluster sample
selection, the \xray $M-T_X$ relation,
the ICM mass measurement, 
the total NIR luminosity, the mean NIR stellar \mlre, and the ICM and stellar
metallicities. These affect the derived NIR luminosity, the stellar
mass fraction, the cluster baryon fraction, and the iron mass fraction, as
a function of cluster mass. In what follows we discuss the robustness of these
ingredients.

\emph{Cluster sample selection:} Our cluster sample is taken partly from an 
\xray flux limited sample and partly from a medley of clusters with 
available \xray temperature measurements.  Our results may depend on this 
selection, because low mass clusters in our sample tend to be at lower 
redshift than the higher mass clusters.  Any redshift related systematic 
could then potentially masquerade as a mass trend.  We examine this 
possibility in Fig~\ref{fig:lstar}, where we compare the luminosity 
function parameters for different mass and redshift ranges.  
As discussed in \S\ref{sec:lf}, after dividing the sample into two mass subsamples, we find
no redshift trends in luminosity function parameters.

\emph{$M-T_X$ relation:} The $M-T_X$ determines the cluster mass and the virial
radius.  We adopt the observed relation based on ASCA measurements of cluster
temperature profiles \citep{finoguenov01}. The self-similar cluster
evolution scenario predicts the slope to be $1.5$, which is
consistent with the observed value we use. Using a slope of $1.5$ (with a
normalization chosen to be the same as the observed one at $T_X = 6$~keV)
produces best-fit relations that are statistically consistent with the fiducial
results. However, we notice that shifts in normalization can be important --
in most of the cases crucial -- to estimates of the cluster baryon fraction,
\mlre, and stellar mass fraction; however, normalization uncertainties do not
affect the trends we see with cluster virial mass.

\emph{ICM mass:} All our ICM masses are from ROSAT PSPC \xray surface brightness
profiles that extend to or near $r_{500}$, based on the adopted $M-T_X$ relation. We 
also take into account the clumping in the ICM and depletion of the baryon fraction
at $r_{500}$ using results from hydrodynamical simulations (MME).

\emph{NIR luminosity:} We use the second release data from 2MASS, which has accurate
photometry and a small star-galaxy misidentification rate.
Major factors that affect the estimates of the total NIR luminosity include: the cluster
angular size (mainly by $T_X$ \& $M-T_X$ relation), the faint-end slope $\alpha$ 
of the luminosity function, the contributions from the BCGs, and the method of
background subtraction. In the case that mass and light have the same
distribution, whether we choose $r_{500}$ or $r_{200}$ should not affect our
results. Currently we are restricted by the sky coverage of 2MASS second release
data: we have NIR data only out to $r_{500}$ for some of the clusters, which
increases the uncertainties in estimating light at $r_{200}$.  We have examined the light--mass relation at $r_{200}$, and the best-fit slope is statistically consistent with that at $r_{500}$.
The faint-end slope of the luminosity function $\alpha$ has very little effect for those 
nearby clusters that 2MASS has probed deeply enough to measure a large fraction of light, but for clusters with larger redshifts, the effects are larger.
Overall, though, the best-fit slopes calculated for several different values of $\alpha$ are consistent with the fiducial ones. 
Finally, the statistical and ``annulus'' methods for
background subtraction give statistically consistent results. It is interesting 
to notice that, based on the observed $K$-band $\log N - \log S$ relation, the 
probability that a foreground galaxy 2 magnitudes brighter 
than $M_*$ lies within the cluster region increases slightly with redshift for
our sample; light from a chance superposition like this is subtracted statistically.  However, if there were any residual bias, the effect would be to add more light to higher redshift (which tend to be more massive) clusters, which would weaken the trend in \mlr that  we observe.

\emph{Mean stellar \mlre:} We use this to estimate the
stellar mass and the associated quantities. Instead of using
a varying \mlr based on the (somewhat uncertain) spiral fraction
and the (also uncertain) stellar \mlr in elliptical and spiral
galaxies, we could assume a constant \mlr given by 2dF survey \citep[][see \S\ref{sec:totlite}]{cole01}.
Using this constant \mlr gives scaling relations that have slopes about $0.5-1\sigma$ different
from our fiducial ones. 

\emph{Mean ICM \& stellar metallicities:} These quantities are used to
convert stellar and ICM mass into iron mass. The ICM iron abundance is measured
to be relatively constant against $T_X$ (De Grandi 2003, private 
communication), and the value that we adopt is representative of the ICM as a
whole. The stellar iron abundance, on the other hand, suffers 
difficulties that are similar to the mean stellar \mlre, due to uncertainties in observations 
and the
relative spiral and elliptical abundance. If we assume solar abundance for stars
in all galaxies, we obtain a $y_{Fe}-M$ relation with slope $\sim 1.5\,\sigma$
steeper than the fiducial one.

In summary, as far as the best-fit results are concerned, all the 
effects considered above give consistent results with our 
fiducial results; the factors that give largest deviations are the 
faint-end slope of the luminosity function and the choice of cluster region.  
Other factors, such as the assumed cosmology in calculating the distance to
the clusters, or the Malmquist bias (less than $3\%$), do not affect our results.
We therefore conclude that the results presented in the 
previous sections are robust.
We look forward to better examining sample selection issues once the full 2MASS dataset is 
available.

\section{Summary \& Prospects}
\label{sec:sum}

We have used the \nir data from the 2MASS survey together with
archival and published \xray data to analyze the properties of the
stellar and ICM baryon reservoirs in galaxy clusters.  
While \xray temperature gives us reliable cluster binding mass
estimates, \nir light traces stellar population better than 
optical bands; the joint analysis makes it possible to examine
whether stellar \nir luminosity can serve as a good cluster mass
estimator. It also enables us to probe the star formation efficiency
for clusters spanning an order of magnitude in mass.
For a subsample of 13 clusters, we also measure ICM
masses from \xray data. We are therefore able to investigate the
relative abundance of the main baryonic components in clusters, as
well as the issues concerning the ICM iron enrichment processes.
We review our conclusions here, and summarize the scaling relations
in Table 3.  

First, the \nir properties of galaxy clusters appear to exhibit
significant regularity.  The $K$-band luminosity within $r_{500}$ is
strongly correlated with the cluster binding mass, with an observed
scatter of approximately 28\% in $K$-band luminosity at a fixed
cluster mass.  Thus, the \nir cluster light is a good predictor of
cluster mass.  The $K$-band \mlr increases from $\Upsilon_{500}\sim33\,h_{70}\,
\Upsilon_\odot$
for $10^{14}h_{70}^{-1}M\odot$ clusters to $\Upsilon_{500}\sim 68\,h_{70}\,
\Upsilon_\odot$ for
$10^{15}h_{70}^{-1}M\odot$ clusters.  This trend must be accounted for
when interpreting \nir cluster surveys, where the \nir light is the
primary indicator of cluster mass. However, we also notice that for clusters
more massive than $2\times 10^{14}h_{70}^{-1}M\odot$ ($T_X\ge 3.7$~keV) there is
little evidence for a changing \mlre. 

Second, the cluster baryon fraction together with constraints from the
CMB observations on $\Omega_b$ provides a measure 
of the matter density parameter $\Omega_M=0.28\pm0.03$ (for $h_{70}=1$) that is in excellent agreement with recent CMB anisotropy results.

Third, the \mlr of the most massive clusters can be used with a
measure of the $K$-band luminosity density of the universe to estimate
the matter density parameter $\Omega_M=0.19\pm0.03$.  Alternatively,
adopting the baryon fraction constraint on $\Omega_M$, we show that
the $K$-band \mlr in the universe must be
$\overline{\Upsilon}_{univ}=(78\pm 13)h_{70}\Upsilon_\odot$.  This \mlr is
marginally consistent with that for our more massive clusters
$\overline{\Upsilon}_{500,hot}=(53\pm3)h_{70}\Upsilon_\odot$, indicating that
differences in the star formation history in these 
environments produce present epoch \mlrs
that differ by as much as $\sim$30\%.

Fourth, the amount of $K$-band light per unit cluster binding mass is a
factor of $\sim2$ times higher for low mass clusters than for high
mass clusters.  Accounting for a slight variation in the mean $K$-band
stellar \mlr for galaxies in low and high mass
clusters, we find that the overall star formation
efficiency decreases with increasing cluster mass.  This observation
provides a new constraint on developing galaxy formation
models.

Fifth, the decreasing stellar mass fraction and increasing ICM mass
fraction lead to a ratio of ICM mass to stellar mass that varies from
5.9 to 10.4 over the cluster mass range of $10^{14}h_{70}^{-1}M_\odot$ to
$10^{15}h_{70}^{-1}M_\odot$.  Thus, stars constitute roughly 14\% of the baryons
in low mass clusters and only 9\% in high mass clusters. 

Sixth,  the trends in stellar mass and ICM mass with cluster binding
mass provide useful constraints on the thermodynamic history of galaxy
clusters.  Specifically, pure preheating models are inconsistent with
our observations, because preheating of the ICM would not introduce a
decreasing star formation efficiency.  In addition, simple
cooling--oriented models that suggest a constant baryon fraction with
a larger portion of the baryons going into stars in low mass clusters
are inconsistent with our observations, because the total baryon
fraction is increasing with cluster mass in our sample.  Models that
include varying star formation efficiency {\it and} preheating or
cooling to achieve a minimum entropy level in the ICM of
$\sim100$\,keV\,cm$^2$ would likely reproduce the observed trends.

Seventh, the iron yield per unit stellar mass is large and an increasing
function of cluster binding mass.  It is likely that unusually
top--heavy stellar initial mass functions would be required to enrich
the baryons to such a high level.  Interestingly, the iron fraction
(ratio of iron mass to cluster binding mass) is roughly constant.
Together with the trend of decreasing stellar mass fraction, this
constant iron mass fraction suggests that there may have been an early
epoch of star formation and enrichment -- perhaps before the epoch of
galaxy formation -- that acted uniformly over the full range of cluster
masses.

Due to the incomplete sky coverage of the 2MASS second incremental
release, our analysis is restricted to a small sample of 27 clusters; the 
shallow 2MASS photometry limits our study to low redshift clusters. We plan to
expand the analysis presented here to a much larger  sample  of local clusters, 
as well as to clusters at higher redshift using deeper photometry.  Together, these studies will lead us
toward a more complete understanding of the evolution of galaxy 
clusters and their baryon reservoirs.  This understanding will 
undoubtedly contribute to interesting cosmological studies and an improved 
understanding of the star formation history of the universe.

\acknowledgements

We thank an anonymous referee for comments that improved the paper.
We extend out thanks to Chris Kochanek, Martin White, Thomas Reiprich and 
Anthony Gonzalez for helpful comments on an early version of the manuscript. 
YTL thanks Subha Majumdar, Al Sanderson and I.H. for helpful discussions.
JJM acknowledges financial support from the NASA Long Term Space
Astrophysics grant NAG 5-11415.  SAS acknowledges financial support from the NASA Long Term Space Astrophysics grant NAG 5-8430. 
This publication makes
use of data products from the Two Micron All Sky Survey, which is a
joint project of the University of Massachusetts and the Infrared
Processing and Analysis Center, funded by the National Aeronautics and
Space Administration and the National Science Foundation.  This
research has made use of the NASA/IPAC Extragalactic Database (NED)
which is operated by the Jet Propulsion Laboratory, California
Institute of Technology, under contract with the National Aeronautics
and Space Administration.

\appendix

Here we describe our estimate of the mean stellar \mlr for each
cluster. As a simplified model, we divide galaxies into 
early and late types and calculate the average
stellar \mlr in the two types. We employ published estimates of the stellar 
\mlr as a function of mass and luminosity in both ellipticals and spirals.  
With published estimates of the spiral fraction as a function of cluster mass 
and of the luminosity function for ellipticals and spirals, we then calculate 
the luminosity-weighted mean stellar \mlre. We describe each of these steps
in turn.

For the $K$-band stellar \mlr in ellipticals, we use the results of 
a dynamical analysis of 21 luminous ellipticals \citep{gerhard01}.
Specifically, from their Fig~13 we estimate the central
\mlr $\Upsilon_{e}(L)$ as a function of galaxy luminosity, using the
typical color for ellipticals $B-K \approx 4.1$ \citep{pahre99}. 
The central \mlr is a good estimate of the stellar \mlre, because the 
central regions of ellipticals are dominated by stellar mass.  For 
the $K$-band stellar \mlr of spiral 
galaxies, we use the results of \citet[see their Fig~1]{bell01},
who construct models that describe many characteristics of
spirals in the Ursa Major cluster.

We adopt the $K$-band luminosity functions (characterized by the
Schechter parameters $\alpha$ and $L_*$) from a $K$-band 2MASS study 
\citep{kochanek01}. With subscripts $e$ \& $s$ denoting ``elliptical' \& 
``spiral'', respectively, the mean stellar \mlr is
\[
\overline{\Upsilon}_{star} = { n_e\,L_{*,e}\, \int_{y_{low,e}}^\infty 
  \Upsilon_{e}(y)\, {\rm e}^{-y} y^{\alpha_e+1} d y
  + n_s\,L_{*,s}\,\int_{y_{low,s}}^\infty \Upsilon_{s}(y)
  \, {\rm e}^{-y} y^{\alpha_s+1} d y
  \over n_e\,L_{*,e}\,\int_{y_{low,e}}^\infty \, {\rm e}^{-y} y^{\alpha_e+1} d y
  + n_s\,L_{*,s}\,\int_{y_{low,s}}^\infty \,{\rm e}^{-y} y^{\alpha_s+1} d y},
\]
where $y_{low,i} \equiv L_{low}/L_{*,i}$ ($i=e,s$),
$L_{low}$ corresponds to $M_{low}=-20$,
the luminosity cutoff (see \S\ref{sec:totlite}),
the galaxy number densities $n_e$ \& $n_s$ are determined from
the total number of galaxies observed
$n_{tot}'\! =\! \int_{L_{low}/L_*}^{\infty}\!\phi_*\,(L/L_*)^{\alpha}\,{\rm exp}(-L/L _*)d(L/L_*)$
and the spiral galaxy number fraction $f_s$:
$n_{e} = n_{tot}' (1-f_s)/\int_{y_{low,e}}^\infty {\rm e}^{-y} y^{\alpha_e} dy$,
$n_{s} = n_{tot}' f_s/\int_{y_{low,s}}^\infty {\rm e}^{-y} y^{\alpha_s} d y$.
As explained earlier, the integration limits are chosen to avoid integrating 
over the faint-end of the luminosity functions whose shape is unknown
and where the behavior of galaxy $\Upsilon(L)$ is also unknown.

We use early published estimates of the spiral fraction to measure the trend in 
$f_s$ with cluster $T_X$ \citep{bahcall77b,dressler80b,dressler88b}. For the 28 
clusters from these data sets  with published $T_X$, we regard all galaxies 
designated as cD, E and S0 as ellipticals, and we ignore irregulars. The
resulting spiral fraction decreases with cluster temperature. The best-fit to 
the $f_s-T_X$ relation is $\log f_s = -0.5\,\log(T_X/4 {\rm keV})-0.4$. For each
cluster in our sample we calculate $f_s(T_X)$, then use the above expression to 
calculate the mean stellar \mlre.  The resulting $\overline{\Upsilon}_{star}$ 
varies weakly with $T_X$ (from $\sim 0.7 \Upsilon_\odot$ at $2$\,keV to 
$\sim 0.8\Upsilon_\odot$ at $10$\,keV). Curiously, the value at low mass end is 
very close to that obtained by \citet{cole01}, using a \citet{kennicutt83} IMF 
(see Fig~\ref{fig:mlstar}).

Let us point out some caveats for the above approach. First of all,
the galaxy mass--to--light ratios are not solely from observations of 
cluster galaxies (elliptical \mlr are observed but not for galaxies in clusters;
spiral \mlr are from models built from observations of cluster galaxies, see 
above), and the luminosity functions used to calculate the mean \mlr are based
on all the galaxies with $cz>2500$\,km s$^{-1}$, which includes both cluster and
field galaxies \citep{kochanek01}.   We simply assume that these ``field'' 
luminosity functions are applicable to cluster environments. Finally, the spiral
abundances presented in \citet{bahcall77b} and \citet{dressler80b} are based on
visual/blue band observations (which may be larger than the values 
obtained when observed at redder bands), and are not 
restricted to a fixed fraction of virial radius or a fraction of the luminosity 
function comparable to that we use in our 2MASS study; thus, the uncertainty in 
$f_s$ is therefore probably significant and should be examined directly in the 
2MASS data.

Following a similar line of reasoning, we calculate the mean stellar metallicity
based on the observed iron abundance for stars in elliptical and
spiral galaxies $Z_e$ \& $Z_s$:
\[
\overline{Z}_{star,Fe} = { n_e\,L_{*,e}\, \int_{y_{low,e}}^\infty 
  Z_e(y)\,\Upsilon_{e}(y)\, {\rm e}^{-y} y^{\alpha_e+1} d y
  + n_s\,L_{*,s}\,\int_{y_{low,s}}^\infty Z_s(y)\,\Upsilon_{s}(y)
  \, {\rm e}^{-y} y^{\alpha_s+1} d y
  \over n_e\,L_{*,e}\,\int_{y_{low,e}}^\infty \,\Upsilon_{e}(y)\, {\rm e}^{-y} y^{\alpha_e+1} d y
  + n_s\,L_{*,s}\,\int_{y_{low,s}}^\infty \Upsilon_{s}(y)\,{\rm e}^{-y} y^{\alpha_s+1} d y}.
\]
For stellar iron abundance in ellipticals we use the results from
a study of 115 E/S0 galaxies in Coma \citep{jorgensen99}.  This analysis showed
that there is no clear correlation between $[Fe/H]$ and galaxy mass
or luminosity. The median value of $[Fe/H]$ is $\sim 0.1$, which corresponds to 
$Z_e \sim 1.2 Z_{\odot,Fe}$. For spiral metallicity, we examine measurements for
14 spirals in cluster/group environments \citep{terlevich02}.  $[Fe/H]$ of these
spirals does not show a clear correlation with galaxy luminosity; the mean 
$[Fe/H] \sim 0.25$, or $Z_s \sim 1.8 Z_{\odot,Fe}$. Assuming these metallicities
are representative for all spirals and ellipticals, the resultant mean 
metallicity $\overline{Z}_{star,Fe}$ is about $1.45$ at $T_X = 2$ keV and
about $1.27$ at $T_X = 10$ keV.

\bibliographystyle{apj}
\bibliography{cosmology,refs}

\end{document}